\documentclass[10pt]{article}

\usepackage[a4paper, total={6in, 8in}]{geometry}
\usepackage[authoryear,longnamesfirst]{natbib}
 
\usepackage{hyperref}
\usepackage{booktabs}

\usepackage{color}
\usepackage{url}
\usepackage[utf8]{inputenc}
\usepackage[T1]{fontenc}
\usepackage{xspace}
\usepackage{xcolor}

\usepackage{paralist}
\usepackage{amssymb}
\usepackage{subfigure}
\usepackage{amsmath}
\usepackage{graphicx}

\usepackage{amsthm}
\usepackage{algorithm}
\usepackage{algpseudocode}
\usepackage{array}
\usepackage{multirow}
\usepackage{xfrac}
\usepackage{makecell}
\usepackage{placeins}

\newcommand{\aedes}{\emph{Ae. aegypti}\xspace}

\definecolor{lightgray}{rgb}{0.7, 0.7, 0.7}
\definecolor{vlgray}{rgb}{0.9, 0.9, 0.9}

\begin{document}

\title{Acoustic Identification of \aedes Mosquitoes\\using Smartphone Apps and\\Residual Convolutional Neural Networks\footnote{Preprint. This work has been submitted to Elsevier for possible publication.}}              
\author{Kayuã Oleques Paim$^1$, Ricardo Rohweder$^2$,\\Mariana Recamonde-Mendoza$^{3,4}$, Rodrigo Brandão Mansilha$^1$,\\Weverton Cordeiro$^3$\\
\\
\normalsize $^3$Instituto de Informática\\
\normalsize Universidade Federal do Rio Grande do Sul (UFRGS)\\
\\
\normalsize  $^1$ Programa de Pós-Graduação em Engenharia de Software (PPGES)\\
\normalsize Universidade Federal do Pampa - Campus Alegrete (UNIPAMPA)\\
\\
\normalsize $^2$Programa de Pós-Graduação em Genética e Biologia Molecular\\
\normalsize Universidade Federal do Rio Grande do Sul (UFRGS)\\
\\
\normalsize $^4$Núcleo de Bioinformática – Hospital de Clínicas de Porto Alegre (HCPA)\\
\\
\normalsize \normalsize kayuapaim.aluno@unipampa.edu.br (K.O. Paim);\\
\normalsize ricardo.rohweder@ufrgs.br (R. Rohweder);\\
\normalsize mrmendoza@inf.ufrgs.br (M. Recamonde-Mendoza); \\
\normalsize mansilha@unipampa.edu.br (R.B. Mansilha); \\
\normalsize weverton.cordeiro@inf.ufrgs.br (W. Cordeiro)
}
\date{}

\maketitle

\begin{abstract}
In this paper, we advocate in favor of smartphone apps as low-cost, easy-to-deploy solution for raising awareness among the population on the proliferation of \emph{Aedes aegypti} mosquitoes. Nevertheless, devising such a smartphone app is challenging, for many reasons, including the required maturity level of techniques for identifying mosquitoes based on features that can be captured using smartphone resources. In this paper, we identify a set of (non-exhaustive) requirements that smartphone apps must meet to become an effective tooling in the fight against \emph{Ae. aegypti}, and advance the state-of-the-art with (i) a residual convolutional neural network for classifying \emph{Ae. aegypti} mosquitoes from their wingbeat sound, (ii) a methodology for reducing the influence of background noise in the classification process, and (iii) a dataset for benchmarking solutions for detecting \emph{Ae. aegypti} mosquitoes from wingbeat sound recordings. From the analysis of accuracy and recall, we provide evidence that convolutional neural networks have potential as a cornerstone for tracking mosquito apps for smartphones. 
\end{abstract}

\section{Introduction}

Mosquito-borne diseases remain one of the most significant health problems worldwide, with estimates of 700 million people contracting these illnesses annually and more than one million killed per year \citep{zika-virus-disease-2018}. In economic terms, these diseases generate severe losses to the global economy, estimated in billions of dollars each year \citep{bradshaw2016massive,diagne2021high,ahmed2022managing}. Among these mosquitoes, we highlight \emph{Aedes aegypti}, which can transmit diseases like Dengue, Zika, Chikungunya, and Yellow Fever.

Fighting the spread of mosquito-borne diseases requires tackling the proliferation of their vector populations. Existing techniques to this end, such as source reduction \citep{forsyth2020source} and mosquito traps \citep{leandro2022citywide}, can be considered locally efficient. However, the adoption of such techniques faces substantial challenges -- whether social, technical, or economic -- to scale in the large. One could mitigate these challenges by monitoring the spread of mosquito populations and their proliferation rates. On one hand, information on mosquito prevalence in a given region may help guide public investment toward areas with higher proliferation. On the other hand, monitoring can also optimize the adoption of expensive techniques like spreading transgenic sterile mosquitoes \citep{waltz2021first}. In summary, carrying out comprehensive and accurate monitoring of mosquito proliferation on spatial and temporal scales is vital to controlling (and even eradicating) mosquitoes in a given region.

Nevertheless, monitoring mosquito proliferation is also challenging, regardless of the sensing technology used (computer vision, audio recognition, etc.) \citep{fernandes2021detecting}. Solutions based on special purpose devices like mosquito traps often have a non-negligible Total Cost of Acquisition (TCA) and Total Cost of Ownership (TCO) \citep{townson2005exploiting}. To reduce these costs, complementary monitoring methods based on crowd-sourcing, using widely available general-purpose devices (such as smartphones), are desirable \citep{mukundarajan2017using,fernandes2021detecting}.

In our previous work \citep{fernandes2021detecting}, we advocated in favor of using smartphone apps for monitoring \aedes proliferation through crowd-sourcing and advanced the state-of-the-art by proposing convolutional neural networks (CNNs) architectures for the classification of audio of mosquito wingbeat recorded with smartphone apps. In particular, we explored three distinct CNN architectures -- binary, multiclass, and ensemble of binary architectures -- and used them to identify a mosquito wingbeat sound in audio captured from smartphone microphones. In that work, we provided evidence that one can obtain better precision and accuracy in acoustic-based mosquito identification using the proposed architectures without resorting to additional metadata (e.g., geospatial data).

From our continued experience in mosquito audio classification using those architectures and recorded using off-the-shelf smartphones, in this paper we conjecture that an effective smartphone app for acoustic mosquito identification (mosquito tracking app) must meet the following (non-exhaustive) set of requirements:
\begin{enumerate}
 \item[(R1)] The mosquito tracking app must run effectively and efficiently on popular off-the-shelf smartphones;
 \item[(R2)] The neural network running in the app must be robust to a wide range of background noise; and 
 \item[(R3)] The neural network must be resilient to bogus noise deliberately generated to induce false positives.
\end{enumerate}

In this paper, we address requirements R1 and R2 by making three contributions. First, we propose a residual CNN architecture to enable effective and efficient mosquito wingbeat classification through audio analysis in off-the-shelf, popular smartphones. Our architecture performs mosquito classification from spectograms of audio segments, using a set of residual and classification blocks to reduce the computing overhead (and, therefore, smartphone battery use) and dropout block to avoid overfitting. The resulting architecture is thus more efficient than existing architectures, also delivering higher precision and accuracy in mosquito classification, thereby addressing requirement R1.

As a second contribution of this paper, we present an original dataset of mosquito wingbeat audio recordings that the research community can use as a benchmark for evaluation of machine learning models for mosquito audio classification. Finally, our third contribution in this paper is a training strategy for reducing the influence of environmental noise on mosquito wingbeat classification, thereby addressing Requirement R2. We also discuss strategies related to noise reduction and noise supression, thereby providing research directions for tackling Requirement R3.

The results of an evaluation made with mosquito wingbeat recordings (including the novel recording dataset and existing ones) provide evidence of the effectiveness, efficiency, and robustness of the proposed architecture in contrast to existing ones. The proposed neural network is 18.5\% smaller in terms of the number of parameters, which is the main computational efficiency factor of the neural network heuristic. The neural network combined with a novel training technique can generate results equivalent to or superior to the state-of-the-art in accuracy, precision, recall, and F1. In addition, implementing a functional prototype provides evidence of the feasibility of running our architecture on a Asus Zenfone Max M3 Qualcomm Snapdragon 430 64-bit Octa core 4G RAM Android 8.1 and a Motorola Moto E6 Play Cortex-A53 1.5 GHz Quad Core 2 GB RAM Android 9.0 smartphones.

The remainder of this paper is organized as follows. In Section~\ref{sec_background} we review technical background and related work. In Section~\ref{sec:methodology} we detail the methodological processes carried out in our study, especially the method for obtaining the recording datasets. In Section~\ref{sec:proposal} we present our neural network architecture and training method to enable mosquito classification in popular, off-the-shelf smartphones. In Section~\ref{sec_eva}, we evaluate and present a proof-of-concept smartphone app. We then close the paper in Section~\ref{sec:conclusion} with concluding remarks and directions for future research.

\section{Background and Related Work}
\label{sec_background}

Strategies for automated mosquito identification based on wingbeat frequency analysis have been studied since 1945~\citep{kahn1945recording}. 
For instance, work involving the use of audio recordings captured by acoustic devices for the identification of West African mosquito species date as early as 1947~\citep{kahn1947}. Nevertheless, the difficulty of collecting audio data with the available technology at the time left researchers capable of obtaining mosquito wingbeat recordings in controlled environments (e.g., laboratory) only~\citep{chen2014flying}, which in turn influenced the quality of the classification models' results.

To avoid such audio technology limitations, researchers have investigated alternatives like photosensors, trap optimization, and computational vision for mosquito classification \citep{li2005automated,potamitis2015insect,silva2015exploring, ouyang2015mosquito, potamitis2016measuring, KimDongmin2021}.
For example, \cite{li2005automated} used photosensors' based dataset to classify five species of mosquitoes, including \aedes, in a controlled environment and obtained an accuracy of 72.67\% using artificial neural networks. More recently, \cite{ouyang2015mosquito} used infrared sensors to record and classify three different mosquito species at a sampling rate of 5,000Hz, and achieved an accuracy of up to 85.4\% with a machine learning classifier (Gaussian mixture model trained using the expectation-maximization algorithm, EM-GMM). Similarly, some studies have explored the potential of using photosensors in passive mosquito traps \citep{silva2013applying, chen2014flying, potamitis2016measuring}.

Several studies have also proposed ways to improve mosquito recording using acoustic and visual stimuli \citep{johnson2016siren, balestrino2016sound, pantoja2019new, staunton2020low}. 
\cite{johnson2016siren} investigated how to use male flight tones to attract female mosquitoes and thus improve their recording using non-mechanical passive Gravid Aedes Traps (GAT). In the same direction, \cite{rohde2019waterproof} used low-cost microcontrollers to generate frequency tones close to that emitted by female \aedes mosquitoes for use in water-resistant and low-cost GAT traps. \cite{balestrino2016sound} followed a similar approach and identified that \emph{Aedes albopictus} mosquitoes are more responsive to acoustic stimulation up to 4 days of age and that the black color of the trap positively influences capture. In the same direction, \cite{villarreal2017impact} analyzed the variation in mosquito wing flapping according to the environmental condition. They identified that the fundamental female frequency highly depends on the ambient temperature, increasing ~8-13 Hz with each °C gain. ~\cite{cator2009harmonic} further found that \aedes mosquitoes change their wingbeat pitch to match during mating. 

In another direction, various authors used computer vision for mosquito recognition and tracking tasks \citep{motta2019, Amiruddin2020, Akter2021,Rakhmatulin2021, Adhane2022}. For instance, \cite{motta2019} implemented a computational model based on a convolutional neural network (CNN) to extract features from mosquito images to identify adult mosquitoes of the species \aedes, \emph{Aedes albopictus}, and mosquitoes of the genus \emph{Culex}. More recently, with the advancement of technology and the quality of audio capture devices, the research community has again focused on traps based on acoustic sensors. ~\cite{vasconcelos2019locomobis} proposed using this type of sensor in GAT traps to capture the sound of flapping wings produced by mosquitoes in flight for automatic identification. However, trap-based approaches remain expensive for monitoring large areas requiring thousands of devices for adequate coverage. Conversely, devices with capture and processing capabilities, such as smartphones, have become ubiquitous in recent years. \cite{mukundarajan2017using} demonstrated that smartphones could be powerful tools for recording mosquito wingbeats, even those with the most basic functionalities.

In a prior work, we investigated neural networks to recognize smartphone-recorded mosquito audio~\citep{FERNANDES2021}. We used Mel scale-based audio spectrograms and a convolutional neural network to classify 22 mosquito species, including \aedes. In that work, we achieved an accuracy of 94.5\% and 78.12\% for multiclass classification considering the species individually and for a binary classification considering two groups: \aedes vs other species. In the same direction, investigations using deep neural networks tried to improve classification results~\citep{kiskin2020bioacoustic, Wei2022}. For example, \cite{Wei2022} used convolutional networks interspersed with self-attention layers to classify a sample set of mosquito audio samples. In another direction, attempts to obtain computationally lighter models were also explored \citep{Yin2021, Hernan2021a, Toledo2021,Alar2021b, Su2022}. \cite{Su2022} proposed a technique involving audio sample rate reduction and lighter classification algorithms.

In order to achieve smartphone apps capable of mosquito classification in the wild, we must address many practical challenges related to audio classification in non-controlled environments using off-the-shelf, popular smartphone hardware.
Among these challenges is the amount of ambient noise, which can significantly affect the classification of sound events \citep{Choi2018} and the limited battery capabilities of the mobile device \citep{Zhao2022}. Some proposals were made to reduce ambient noise and mitigate the impact of noise on the classifier. Regarding audio enhancement, \cite{Salih2017} used low-pass filters to reduce noise interference in audio quality. Regarding the classifier's quality, \cite{Yin2015} proposed the addition of random noise in the audio samples during the training stage to make the classifier more resilient to noise. In this paper, we explore these directions to improve precision and accuracy in mosquito wingbeat classification from audio recordings by means of a lightweight convolutional neural network.

\section{Materials and Methods}
\label{sec:methodology}

In this section, we present the materials and methods used to construct our solution based on deep learning. We start with Subsection~\ref{sec:methodology:data} describing the audio sample collection process and detailing the available data set. 
In  Subsection~\ref{sec:methodology:features}, we present the audio processing and feature extraction methods used and discuss aspects related to the characteristics and peculiarities of the processed audios. 
We describe the proposed neural network topology in Subsection~\ref{sec:methodology:neural}. Finally, we define an evaluation model in  Subsection~\ref{sec:methodology:model_evaluation}.

\subsection{Collecting Mosquito Audio Recordings} 
\label{sec:methodology:data}

The datasets of raw audio files analyzed in this work can be organized into three sets, as detailed in Table~\ref{tab_raw_audio}: (i) recording samples containing no mosquitoes; (ii) recording samples containing mosquitoes that have been previously investigated; (iii) recording samples containing novel mosquito recordings. The first set helps to qualify the training of neural networks and to avoid bias in the evaluation process. The second set facilitates the comparison of the proposed solution with previous investigations in the field. The third set allows us to advance the state of the art with greater flexibility, as it was conceived, planned, and executed following specific research objectives.
  
The first set (Raw recording set RD1) comprises recording samples from environments without mosquitoes and with different noise sources and intensity levels. These samples make the neural network more resilient to interference caused by environmental noise. They also balance the assessment fairly and thus avoid benefiting biased networks (e.g., a network trained to look for a \emph{needle in a pile of needles}). To achieve these goals, we used the ESC-50 audio set \citep{piczak2015dataset}, which is publicly available and has been used in several studies \citep{AST21, conectionDot22}. This set has 2000 audio samples from 40 different noise classes. Each audio sample lasts 5 seconds and has a sampling rate of 44.1kHz.

For the second set of recordings (RD2, RD3, and RD4), we adopted the data generated by the Abuzz project \citep{mukundarajan2017using}. This set consists of 1,285 audio files, which include 20 species of mosquitoes. Audios available in this set were recorded using commercially available cell phones. In the original work \citep{mukundarajan2017using}, the authors compared eight smartphone models (eg, iPhone 4S and Xperia Z3 Compact). The authors concluded that all the evaluated devices could record the sound emitted by mosquitoes at a distance of up to 50 mm. Most of the sounds captured were from mosquitoes free-flying in various locations, with varying background noise levels. Laboratory recordings of tethered or caged mosquito populations were also performed for comparison. As a result, for example, authors found a significant increase in the degree of overlapping frequencies in records of populations trapped in the laboratory (with little noise) compared to records of free populations flying in different environments.

The third set of audio recording samples (RD5, RD6, RD7, RD8, and RD9) we generated for the purposes of this investigation. We obtained recordings for different genera (e.g., \emph{Aedes}, \emph{Anopheles}), species (e.g., \aedes, \emph{Ae. albopictus}), and sex (i.e., male and female). We performed the mosquito recordings at the insectary of the Federal University of Rio Grande do Sul (UFRGS), Porto Alegre, Brazil, from February 07, 2021 to February 22, 2021, between 08:37 AM and 09:17 PM. Mosquito ages during recording varied between 1 and 12 days. The ambient brightness was maintained at approximately 1,110 lux. The recordings were done using an LG K10 K430TV Octa-core 1.14GHz ARM Cortex-A53 1GB RAM Android 6.0 smartphone. The captured sounds were of free-flying mosquitoes in a cage with sizes between 5 and 10 cubic meters. The method chosen to keep the mosquitoes close to the microphone was using a glass with a shield to prevent further movement of the mosquito. We are publishing those audios so that practitioners can replicate our research, verify our results, and perform further investigation in the field.

\begin{table}[b]
    \centering
\setlength\tabcolsep{7pt}
    \resizebox{1\columnwidth}{!}{%
    
    \begin{tabular}{ l l l c r r r r} 
    
\textbf{Subset} &  \textbf{ID} &    \textbf{Species} & \textbf{Gender} &  \makecell[r]{ Number of\\Mosquitoes } &  \makecell[r]{ Number of\\Recordings}  & \makecell[r]{ Recording Total\\Lenght (s)} \\ 
          \toprule 
          
   No & RD1 &No mosquitoes &	NA & 0	 &	2,000 &	10,000  \\
   
   \midrule  
   
 Classic dataset &  RD2 & \aedes  & F/M & 1 & 22  &  1,736 \\ 
\citep{mukundarajan2017using}& RD3 & \emph{Ae. albopictus}&	F/M   &	1&	7 &	966  \\
&RD4 &Non-\emph{Aedes} &	NA &1 &	871 &	13,237  \\

    \midrule
Novel dataset & RD5 & \aedes  & F & 1 & 192 & 4,607 \\ 
&RD6 & \aedes               & M & 1 &	345	& 8,279 \\
&RD7 & \emph{Ae. albopictus} &	F   &	1&	19&	570 \\
&RD8 & \emph{Ae. albopictus}               &	M	&  1 &  17&	510 \\  
&RD9 &Non-\emph{Aede}s &	NA &	0&	158&	4,740 \\

 \bottomrule
    \end{tabular}
}
    \caption{Raw audio datasets. NA stands for \emph{Not Available}, F stands for \emph{Female}, and M stands for \emph{Male}.}
    \label{tab_raw_audio}
\end{table}

We combined the raw recording sets as described in Table~\ref{tab_segmented_audio} to create various benchmarking datasets. Our goal was to create a representative set of recordings with increasing complexity. For example, Dataset 1 (D1) is adequate to verify the presence or absence of \aedes, while Dataset 5 (D5) enables us evaluating the presence and classification among different species. Furthermore, Dataset 6 (D6) facilitates comparison with our previous work.
  
\begin{table}[b]
    \centering
\setlength\tabcolsep{7pt}
    \resizebox{1\columnwidth}{!}{%
    \begin{tabular}{c l l r r} 
   &  & \textbf{Raw} & \textbf{Number of positive} & \textbf{Number of negative}\\ 
\textbf{ID} & \textbf{Description} & \textbf{Datasets}& \textbf{\aedes} & \textbf{\aedes}\\ \toprule
 
D1 & M \aedes plus No mosquitoes   & \{RD1,RD6\}	& 345	& 2,000 \\
D2 & F \aedes plus No mosquitoes  & \{RD1,RD5\} &	192 & 	2,000\\
D3 & F/M  \aedes plus No mosquitoes    & \{RD1,RD5,RD6\} & 	537 & 	2,000 \\
D4 & F/M \aedes plus Non-\emph{Aede}s mosquitoes &\{RD5,RD6,RD9\} & 537 &	158	 \\	
D5 & F/M \aedes plus No mosquitoes \\  & plus Non-\emph{Aede}s mosquitoes & \{RD1-RD5-RD6-RD9\} & 537 &	2,158 \\ 
D6 & Original dataset \citep{mukundarajan2017using} & \{RD2-RD3-RD4\} & 22 &	878 \\ \bottomrule
    \end{tabular}
}
    \caption{Segmented audio datasets. F stands for \emph{Female} and M stands for \emph{Male}.}
    \label{tab_segmented_audio}
\end{table}

\subsection{Processing Audio to Extract Features}
\label{sec:methodology:features}

\begin{figure}

    \centering
    \includegraphics[width=.95\textwidth]{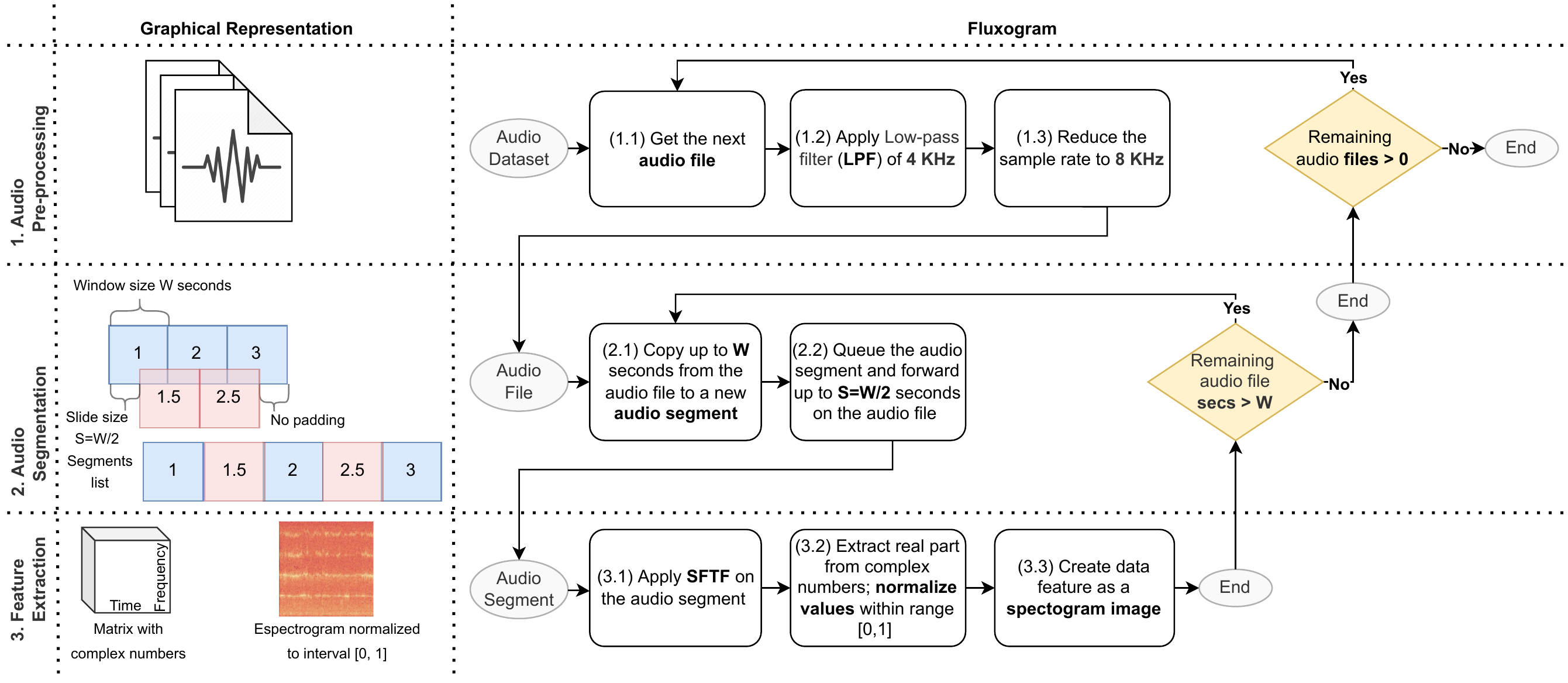}
    \caption{Steps involved in transforming audio datasets into audio features for neural networks.}
    \label{fig_processing}

\end{figure}

Sound representation in the time and frequency domains are helpful as learning resources in audio classification \citep{huzaifah2017comparison}. Works investigating applications of this type of representation to problems related to machine learning, such as audio recognition and speech synthesis, have also been investigated \citep{audioRecognition22, tacotron2}. For these reasons, we adopt such representation in our work. Figure~\ref{fig_processing} illustrates the steps we used to transform audio datasets into features recognizable by neural networks in the time and frequency domains. We organized the process into three main steps: (1) Audio Pre-processing, (2) Audio Segmentation, and (3) Feature Extraction. These steps are detailed in the following paragraphs.

\textbf{Audio Pre-processing}. This step aims to ensure that all audio files are standardized, with the same frequency and sample rate. For each file (step 1.1), we first apply a low-pass filter (1.2) with a frequency threshold of 4 KHz, corresponding to half of the desired sampling rate. This is due to the Nyquist-Shannon Theorem \citep{Shannon1949}, which states that to convert an analog signal into a digital signal that is reliable and viable, the sampling rate must be at least twice as high as the frequency of the signal of interest. Then we reduce the sample rate to 8KHz (1.3). This value is the lowest sampling rate present in the complete set of datasets, including the comparison set (D6) proposed in \cite{mukundarajan2017using} and previously used in  \cite{FERNANDES2021}. We used the Audacity software (version 2.3.3 for Linux) to apply both the low pass filter and downsampling.

\textbf{Audio Segmentation}. This step aims to divide an audio file into several smaller segments with some degree of overlap between them. Overlapping frames is a well know technique adopted to expand the number of samples available for classification and evaluation. As seen in the graphical representation, we generate overlapping frames through a sliding window. More specifically, we assume a given audio file and placeholder that initially points to the beginning of the file and can be advanced. First, we copy an audio segment of size up to \footnote{The file may be shorter than desired.} window size $W$ seconds (2.1).
We then queue up the copied audio segment to be processed in the next step and advance the placeholder of the audio file by slide size $S=W/2$ seconds. We chose this value for the slide size based on our previous study \citep{FERNANDES2021}.
 
\begin{figure}[htb]
\centerline{
    \subfigure[\aedes Male]
    {\includegraphics[width=.21\textwidth]{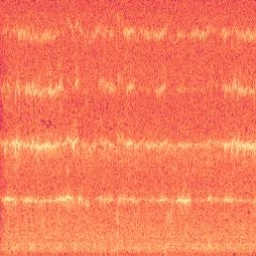}
    \label{fig:spec_aedes_aegypti_male}} 
    
        \hfil
        
    \subfigure[\aedes Female]
    {\includegraphics[width=.21\textwidth]{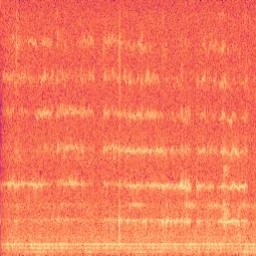} \label{fig:spec_aedes_aegypti_female}} 
    
        \hfil
    \subfigure[\emph{Ae. albopictus} Male]
    {\includegraphics[width=.21\textwidth]{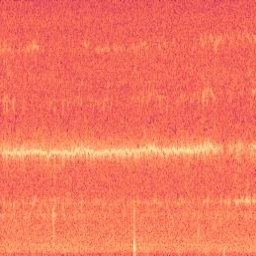}
    \label{fig:spec_aedes_albopictus_male}} 
    
       \hfil
     
    \subfigure[\emph{Ae. albopictus} Female]
    {\includegraphics[width=.21\textwidth]{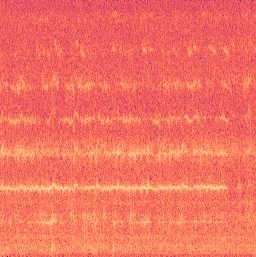}
    \label{fig:spec_aedes_albopictus_female}} 
    
}

\centerline{
    \subfigure[\emph{Ae. sierrensis}]
    {\includegraphics[width=.21\textwidth]{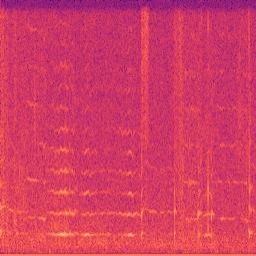}
    \label{fig:spec_aedes_sierrensis}} 
    
    \hfil
    
    \subfigure[\emph{Anopheles quadrimaculatus}]
    {\includegraphics[width=.21\textwidth]{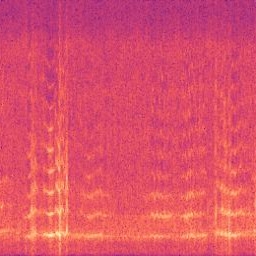}
    \label{fig:spec_anopheles_quadrimaculatus}} 
    
    \hfil
    
    
    
     \subfigure[No mosquitoes]
    {\includegraphics[width=.21\textwidth]{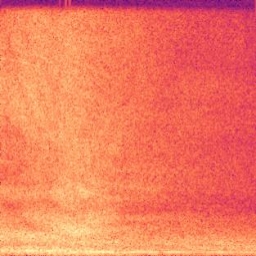}
    \label{fig:spec_environment_noise}} 
    
    \hfil
    
      \subfigure[\aedes Male using prior methodology]
    {\includegraphics[width=.21\textwidth]{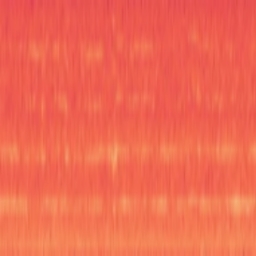}
    \label{fig:spec_prior_work}}} 
   \caption{Spectrograms of audio segments obtained from the D5 dataset generated using proposed (a-g) and prior (h) methodologies.}
   \label{fig_spectograms}
   
\end{figure}

\textbf{Feature Extraction}. This step aims to transform the audio segments into recognizable features for the neural network in the form of an image, as exemplified in Figure~\ref{fig_spectograms}. Each segment/spectrogram/feature must be labeled to facilitate supervised training. The label is defined at a later stage as it varies according to the purpose of the study (e.g., to identify the presence of any mosquito or to identify the presence of a male \aedes mosquito).
 
 First (step 3.1), we decompose the audio segment into a spectrogram by applying the Short-Term Fourier Transform (STFT) to extract features. The results obtained from this decomposition form a matrix with dimensions $m \times n \times 1$. Each element in this matrix refers to a pixel in the spectrogram with the value corresponding to the signal strength in decibels of a given frequency ($m$) at a given point in time ($n$). The data received from the previous step are complex numbers whose real parts can have values in the range of -80 to 0 dB. Given this, we extract only the real part of the complex numbers and normalize the values to the interval [0,1] using  Equation~\ref{equation_norm} (step 3.2), before generating the images (step 3.3).

\begin{equation}
\label{equation_norm}
X_{norm} = \frac{x}{80} + 1
\end{equation}

From Figure~\ref{fig_spectograms}, one can distinguish each of the generated spectrograms visually. For example, recordings containing samples of mosquitoes (Figure~\ref{fig_spectograms}(a-f)) are characterized by the presence of specific patterns of harmonics distributed over several frequency bands of interest -- a phenomenon that does not occur in samples having no mosquitoes (Figure~\ref{fig:spec_environment_noise}). 
We can also identify differences in harmonic distributions for each mosquito specie. For example, the fundamental frequency of the male \aedes mosquito (Figure~\ref{fig:spec_aedes_aegypti_male}) is considerably higher than the male \textit{Ae. albopictus} (Figure~\ref{fig:spec_aedes_albopictus_male}). It is also possible to observe differences by analyzing the sex of the mosquito. Female mosquitoes (Figure~\ref{fig_spectograms}(b,d)) have, in general, a lower frequency of emission and consequently a greater amount of harmonics than male mosquitoes (Figure~\ref{fig_spectograms}(a,c)).
 
Finally, observe that the methodology proposed in this work (e.g., Figure~\ref{fig:spec_aedes_aegypti_male}) generates spectrograms with higher quality than the spectrograms generated by the process proposed in our previous work considering the same audio segment as input (Figure~\ref{fig:spec_prior_work}). In short, with the removal of the Mel scale and other improvements, such as observance of the Shannon-Nyquist theorem, we can identify mosquitoes with closer sound emission ranges.

\subsection{Neural Networks}
\label{sec:methodology:neural}

Our approach to mosquito wingbeat sound classification relies on convolutional neural networks (CNNs). CNNs are deep learning network architectures used in various tasks, including speech and image recognition. An essential feature of this architecture for this application is its ability to recognize patterns and predict results based on input data.

The CNNs are feedforward networks whose core architecture is based on convolutional and pool layers grouped in modules alternately~\citep{10.1162/neco_a_00990}. Convolutional layers are used to extract features from the image by applying trainable convolutional filters while pooling layers progressively reduce the feature size.

Our solution is based on a specific subtype of convolutional neural network based on residual blocks~\citep{residual_2016}. Residual CNNs have shown superior performance over traditional ones. We chose this architecture because it presents significantly better classification performance than pure convolutional models maintaining an equivalent number of parameters.

\subsection{Model Evaluation}
\label{sec:methodology:model_evaluation}

We use the datasets $d \in \{D1, D2, D3, D4, D5\}$ in the evaluation process. For each dataset, we pre-process the respective audio files, generate the spectrograms that serve as input features for the neural networks, and label them according to the question of interest (e.g., is there the presence of \aedes mosquito?). Then, we divided the results randomly into 10-folds and later used them in stages of training and testing the models following the cross-validation strategy.

In the process of comparison with the topology previously proposed by \cite{FERNANDES2021104152}, we tried to reproduce as much as possible the methodology proposed in that work. This includes using the same dataset, which we call $D6$, and the same size and resolution of audio segments. The only change made to the methodology was applied to the dimensions of the input features (i.e., audio pre-processing output) to make data with different formats (e.g., size and resolution) compatible with our neural networks.

For the quantitative evaluation of the results, we used traditional metrics that allow us to observe the results both from the point of view of hits (True Positives and True Negatives) and mistakes (False Positives and False Negatives). Specifically, we use the following metrics.

\begin{itemize}
    \item \emph{Accuracy}: The proportion of instances in the test set that were correctly predicted, divided by the total number of instances in the test set. While precision is a commonly applied and easy-to-interpret metric, it can become an unreliable measure of model performance for class unbalanced datasets, especially for severe distortions in class distributions.
    
    \item \emph{Precision}: The number of correctly predicted positive instances (i.e., TP) divided by all positive predictions made by the model. Low precision indicates a high number of false positives.
     
    \item \emph{Recall}: The number of true positives predicted by the model divided by the total number of positive instances in the test set. The recall is also called Sensitivity or True Positive Rate, and a low recall indicates a high number of false negatives.
      
    \item \emph{F1-measure (F1)}: The harmonic mean between precision and recall, assigning equal weights to false positives and false negatives.

\end{itemize}

\section{Proposed Approach}  
\label{sec:proposal}

Next we describe the proposed solution to the mosquito classification problem through wingbeat recording analysis. We start in Subsection~\ref{sec:proposedApproach:topology} by presenting the proposed neural network topology and discussing its characteristics. 
We then describe in Subsection~\ref{sec:proposedApproach:Training} the training process, list the parameters used, and analyze the model predictions through gradient-weighted class activation mapping.

\subsection{Topology}
\label{sec:proposedApproach:topology}

\begin{figure}
    \centering
    \includegraphics[width=.98\textwidth]{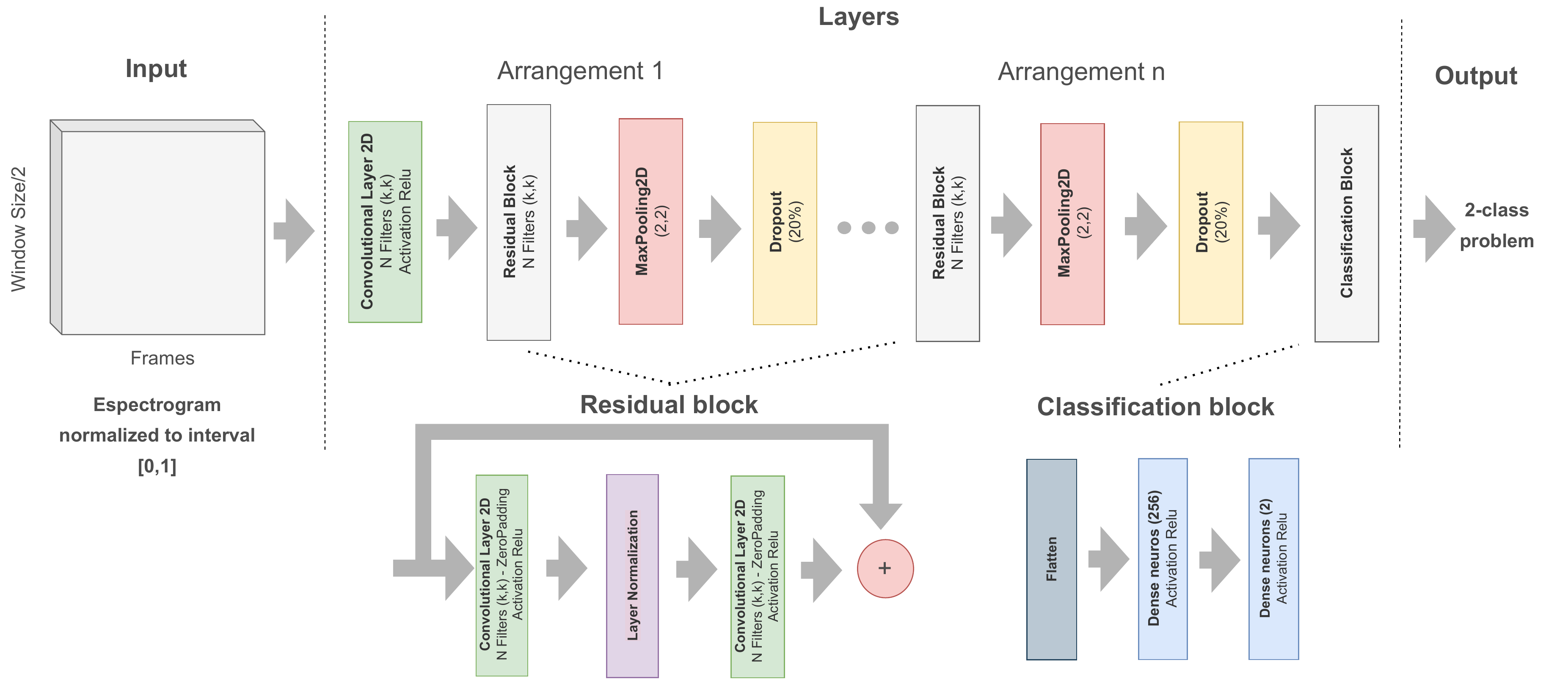}
    \caption{Proposed Neural Network Topology.}
    \label{fig_proposed_nn}
\end{figure}

Figure~\ref{fig_proposed_nn} presents the topology of the proposed neural network architecture. 
The input is formed by a two-dimensional matrix of real values in the interval $[0, 1]$. We chose this configuration due to the format of the spectrograms that form the input features. The output is formed by a vector containing two real values contained in the interval $[0, 1]$ that indicate the probability that the sample represents a sample labeled as true (for example, that contains evidence of an \aedes mosquito\footnote{As will be seen in the next section, on training, we shaped the training specifically for the detection of mosquitoes of the \aedes species, but we could adopt this same topology to identify other species by changing only the sample labels.}).

The intermediate layers are formed by a configurable amount of one or more predefined arrays organized in series and, finally, exactly one classification block. In short, more arrays give a more accurate result but with a higher training and execution cost.
Each array is formed by a sequence of three blocks: residual block, maxPooling2d, and dropout. We chose this configuration to allow the resource to shrink as it traverses each array, significantly reducing the network's computational cost. The use of a dropout layer is configured with a 20\% decay in each arrangement and reduces the possibility of overfitting \citep{dropout2014}.

The \textit{residual block} comprises two two-dimensional convolutional blocks with different filter sizes, Relu activation function, and step equal to (1,1), and a normalization block configured with epsilon=0.001. This configuration is similar to the one used in residual convolutional neural networks \citep{10.1007/978-3-319-46493-0_38, 9879745}. However, we adapted them to reduce the computational cost. Note that there is a connection from the input directly to the output parallel to the sequential flow. The results of the two streams are summed to generate the output values. This configuration, known as residual \citep{10.1007/978-3-319-46493-0_38}, was used to avoid gradient dissipation, which could reduce the model's effectiveness.

The \textit{classification block} transforms the data processed by the neural network into results that are interpretable by the rest of the classification algorithm and consist of three layers. The idea of this configuration is to allow the results in the form of matrices obtained from the previous layer to be interpreted and returned in the form of an interpretable vector. This structure is commonly found in convolutional neural networks, for example, DenseNet \citep{denseNet}. The first is a layer known as flatten, which transforms the matrices from the previous layer into a vector of values. The subsequent two layers contain fully connected neurons with Relu activation function, with the first layer containing 256 neurons and the second having 2 neurons.
 
\subsection{Training}
\label{sec:proposedApproach:Training}

To train the proposed neural network, we label each audio sample in a binary form indicating the presence (true) or absence (false) of \aedes mosquitoes. In other words, in this study we did not try to classify samples according to sex or other species (e.g., \emph{Ae. albopictus}, \emph{Anopheles quadrimaculatus}). For separating the samples and the training and validation set, we used the cross-validation process with 10 folds. In addition, to guarantee the correct balancing of the samples, we defined the probability of selecting a sample proportionally to the quantity of each class (positive or negative). This technique allows for a better balance between classes and helps to avoid possible bias related to sample proportions. Finally, as a loss, we adopt the Binary Crossentropy function as defined in Equation~\ref{loss}. 

\begin{equation}
\label{loss}
\mathrm{Loss} = -\sum_{i=1}^{\mathrm{C}} y_i \cdot \mathrm{log}\; {\hat{y}}_i
\end{equation}

To reduce the influence of environmental noise on the results, we added environmental noise samples obtained from the ESC-50 dataset to the non-\emph{Aedes} mosquito sample set. This is an improvement over our previous work, in which we only considered samples containing mosquitoes (\aedes vs. non-\aedes) in the training process. Although this strategy makes it possible to differentiate \aedes mosquitoes from other species with a high level of accuracy, it tends to fail in an experimentation in the wild, where there is an infinite variety of sounds. This phenomenon should occur more frequently as the presence of sounds with frequencies become closer to \aedes mosquitoes.
 
Another precaution we adopted was to ensure a uniform distribution of samples within each of the classes in the training set. This is because the number of labels is not uniform in the segmented audio datasets (Table~\ref{tab_segmented_audio}). For example, dataset $D1$ contains 345 positive samples and 495 negative samples. An imbalance can degrade the classifying process. To that end, we added an extra step in the cross-validation process, to establish different probabilities of selecting a sample of a given class based on the proportion of similar samples present in the set. In the development process, we use heatmaps based on Gradient-weighted Class Activation Mapping \citep{GRADCAN2017}. In this process, we extract the spectrograms of different positive and negative samples for \aedes mosquitoes and submit them for classification through our pre-trained neural network model.
 
\begin{figure}[htb]

\centerline{

    \subfigure[\aedes Male]{
    \includegraphics[width=.21\textwidth]{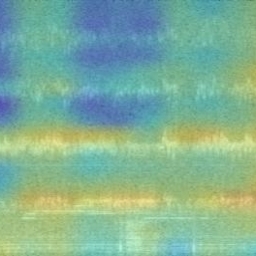}
    \label{fig:gradcam_aedes_aegypti_male}} 
    
    \subfigure[\aedes Female]{
    \includegraphics[width=.21\textwidth]{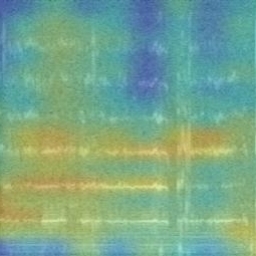} \label{fig:gradcam_aedes_aegypti_female}} 
    
    \subfigure[\emph{Ae. albopictus} Male]{
    \includegraphics[width=.21\textwidth]{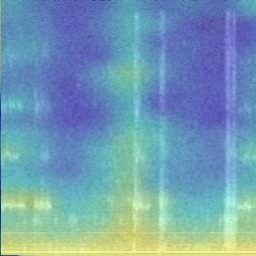}
    \label{fig:gradcam_aedes_albipictus_male}} 

    \subfigure[\emph{Ae. albopictus} Female]{
    \includegraphics[width=.21\textwidth]{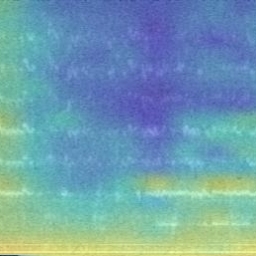} \label{fig:gradcam_aedes_albopictus_female}} 
       
}

\centerline{

    \subfigure[\emph{Aedes sierrensis}]{
    \includegraphics[width=.21\textwidth]{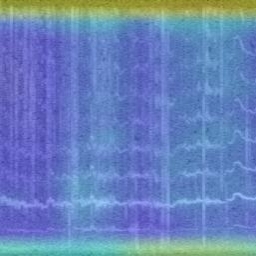}
    \label{fig:gradcam_aedes_sierrensis}} 
    
    \subfigure[\emph{Anopheles quadrimaculatus}]{
    \includegraphics[width=.21\textwidth]{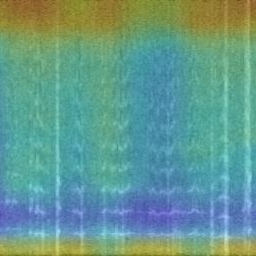} \label{fig:gradcam_anopheles_quadrimaculatus}}

    \subfigure[No mosquitoes]{
    \includegraphics[width=.21\textwidth]{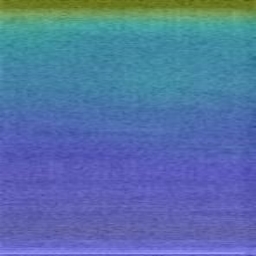}
    \label{fig:gradcam_environment_noise}} 
    
     \subfigure[ \aedes Male using
prior methodology]{
    \includegraphics[width=.21\textwidth]{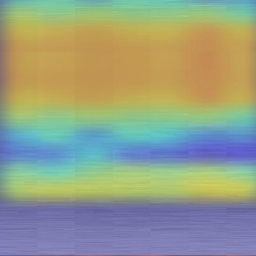}
    \label{fig:gradcam_mel_Aedes_Aegypti}}

}
\caption{Grad-CAM class activation map for different examples}\label{fig_mapa_calor}
\end{figure}

Figure~\ref{fig_mapa_calor} shows activation maps for eight distinct audio samples. The warmer color regions represent areas of greater impact on the class prediction, while the darker areas represent the model's low affinity for the region. From the visual analysis, we can see the high specificity of the model for the patterns present in \aedes positive samples. In particular, greater activation is observed in regions close to the frequencies of 450Hz and 900Hz for positive samples. We also found a greater affinity of the model for the first two harmonics of the sound of male \aedes mosquitoes and for the first three harmonics of female \aedes mosquitoes. We notice proposed solution (Figure~\ref{fig_mapa_calor}(a)) shows more detailed maps than our prior solution (Figure~\ref{fig_mapa_calor}(h)).
  
\section{Experiments and Results}
\label{sec_eva}

We implemented the proposed neural network architecture using Python (v3.7.11). We used Tensorflow (v2.4.1) to implement the topology, Scypy (v1.10.1) and Librosa (v0.10.0) to extract features, and Scikit-learn (v1.1.0) to generate layering 10 -fold cross-validation. In this section, we present an evaluation of our proposal using our proof-of-concept implementation. In Subsection~\ref{subsec_eva_sensitivity}, we present a parametric sensitivity evaluation of the proposed neural network.
In Subsection~\ref{subsec_eva_efficacy}, we present an evaluation of the effectiveness of the proposed neural network in comparison with state-of-the-art solutions. In Subsection~\ref{subsec_eva_proof}, we present a proof of concept demonstrating the feasibility of running the proposed neural network on a low-cost smartphone.

\begin{table}[htb]
    \centering
    \begin{tabular}{l l c l}
  
 & \textbf{Parameter} &	\textbf{Symbol} & \textbf{Values}	\\
 \hline
 
  \multirow{4}{*}{\rotatebox[origin=c]{90}{Input}}  
&Dataset                & $D$ & $\{D1, D2, D3, D4, D5\}$   \\
&Size of Frame          & $F$ & $\{50, 60, 70\}$   \\	
&Size of STFT Window    & $W$ & $\{256, 512, 1024\}$ \\
&Hop Length	            & $H$ & $\{128, 256, 512\}$	   \\
\hline

\multirow{7}{*}{\rotatebox[origin=c]{90}{Neural Network}}
&Size of Kernel Filters     & $K$   &  $\{(3, 3), (5, 5), (7, 7)\}$		 \\ 
&Number of Blocks           & $B$      & 	$\{3, 4, 5\}$			 \\
&Filters Per Block          & $P$      & $\{16, 32, 64\}$ \\
&Sample Rate                &       & 	8 KHz \\
&Size Batch                 &       & 	32 \\
&Feature Type Scale         &       & 	Short Time Fourier Transform \\
&Loss Function              &       & 	Binary Crossentropy \\

\hline	
 
\multirow{6}{*}{\rotatebox[origin=c]{90}{Testbed}} 
& 	Cloud       & & 	OS: Ubuntu 20.04.3 LTS \\
& 	            & & 	CPU: Intel Xeon Gold 6330 3.1GHz \\
& 	            & & 	GPU: Nvidia Tesla V100 16GB \\	
& 	            & & 	RAM: 32GB \\
& 	            & & 	 \\
&  Smartphone   & &     OS: Android v8-10	  \\	
&               & &     CPU: Cortex-A53 1.5GHz Quad Core	  \\
&               & &     RAM: 2 GB	  \\
\bottomrule
    \end{tabular}
    \caption{Evaluation settings.}
    \label{tab_evaluation_settings}
\end{table}

\subsection{Parameter Sensitivity Analysis}
\label{subsec_eva_sensitivity}

We start by evaluating the impact of model parametric values and input features on the results. Table~\ref{tab_evaluation_settings} summarizes the main parameters and respective values considered in the evaluation. As the search space grows exponentially as a function of the number of combinations of parametric values, we adopted two strategies for pruning the branches of possibilities. First, we limit the number of free parameter values to three, for which we try to choose two threshold values, lower and upper, and an intuitively reasonable intermediate value. Second, we used a greedy strategy: we initially arbitrated an intermediate value for each parameter, looking for and fixing the best value for studying the following parameters. We consider the set of kernel filters $K = [(3, 3), (5, 5), (7, 7)]$ because they are widely used values in the literature. Regarding the impact of the number of blocks, we consider the set $B = [3, 4, 5, 6]$ as they are values in the range for which the number of parameters and filters of the neural network remains low, which favors performance. As for the frame size, we consider the set $F = \{50, 60, 70\}$, as these values are equivalent to time intervals between 2.5 and 5.0 seconds of recording. For the STFT window size, we evaluated the set of values $W = \{512, 1024, 2048\}$ since they are values commonly used for audios with a sampling rate below 8kHz. For the number of filters per block $P=\{16, 32, 64 \}$ and hop length $H=\{128, 256, 512\}$, we tried to verify relatively small values to simplify the processing but generated satisfactory results.

\begin{figure}[htb]
\centerline{

    \subfigure[Size filters \\ (B=4, F=60, W=512, P=32, H=256)]{
    \includegraphics[width=.30\textwidth]{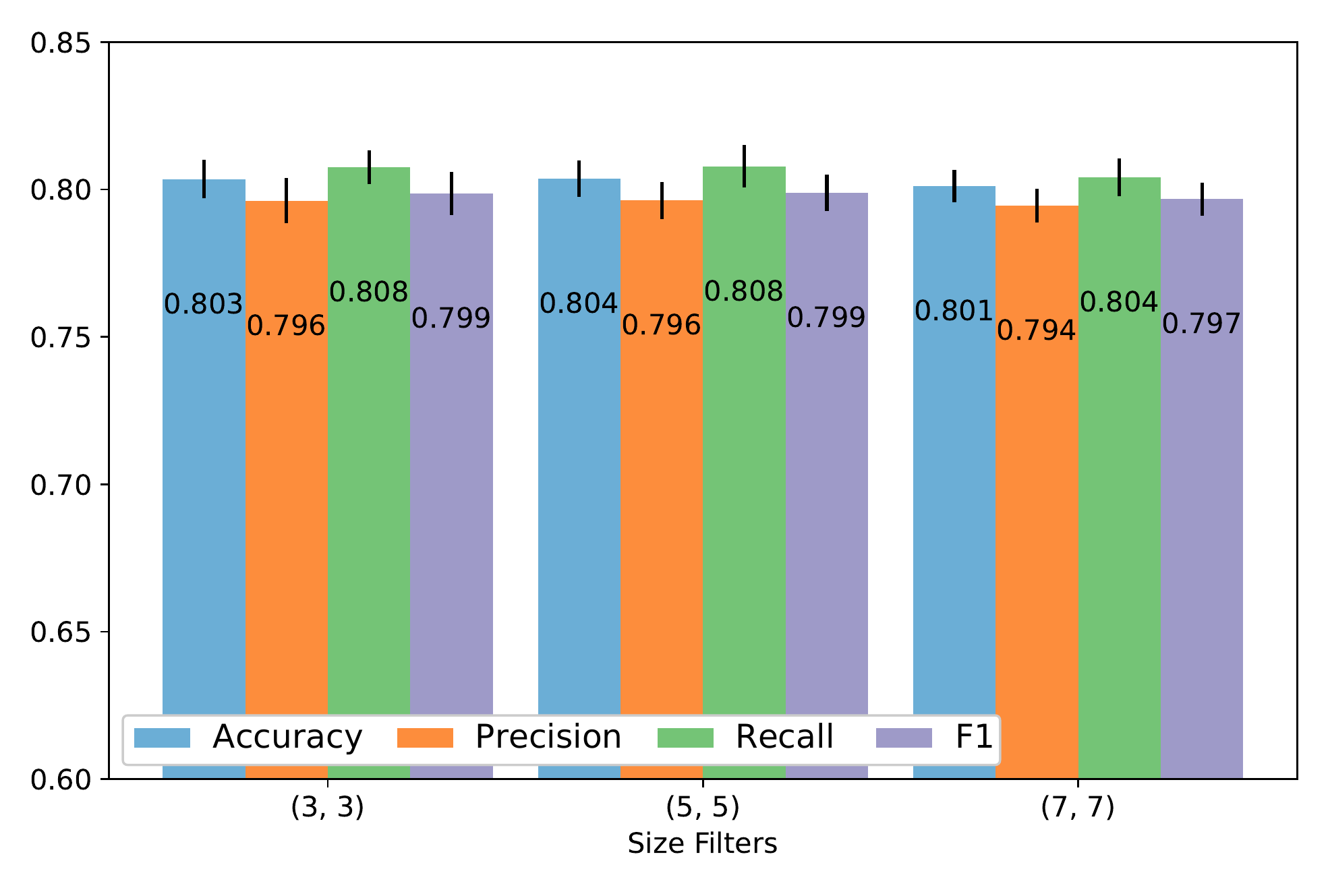}
    \label{fig:size_filters}} 
    
    \subfigure[Number blocks \\ (K=(5, 5), F=60, W=512, P=32, H=256)]{
    \includegraphics[width=.30\textwidth]{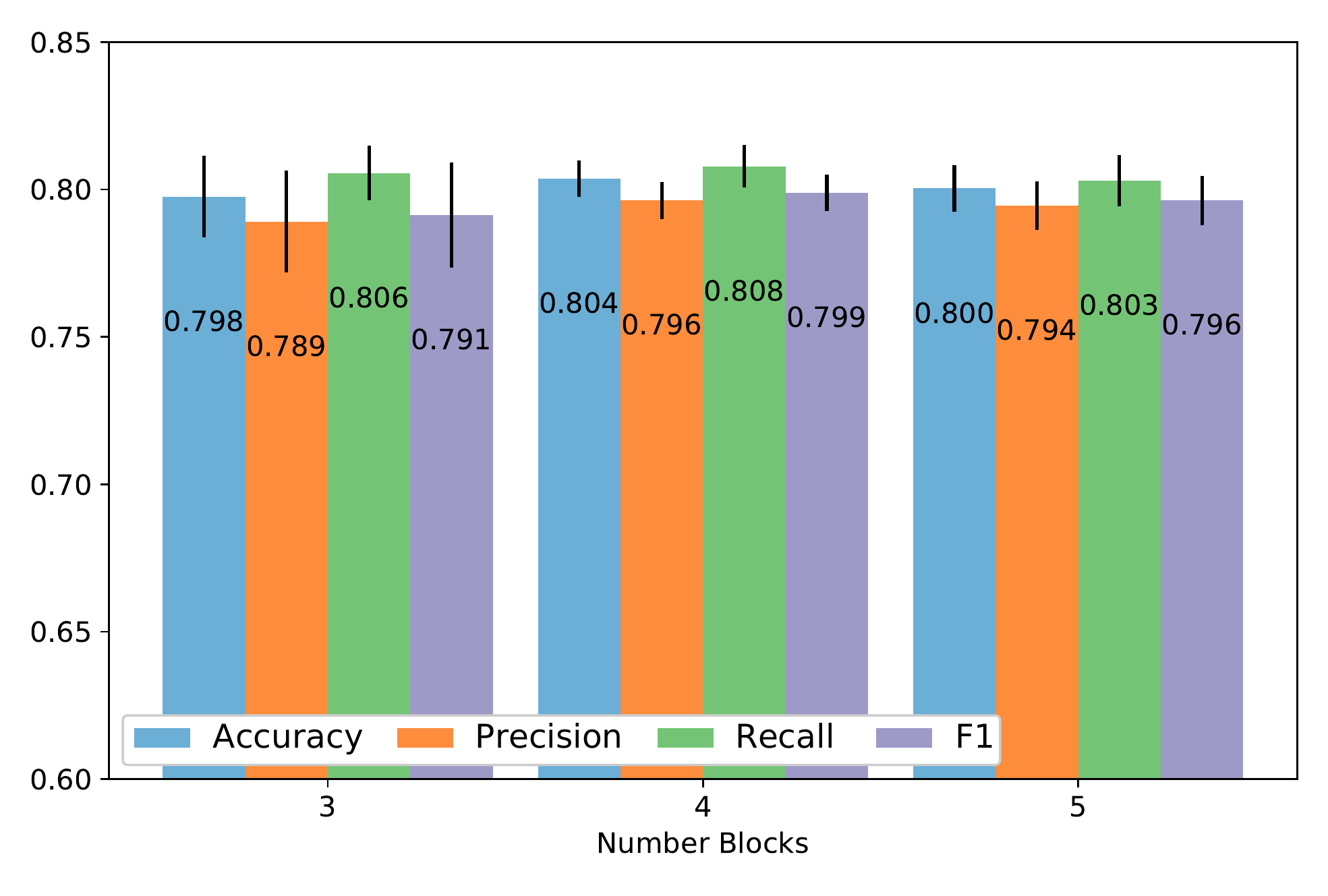} \label{fig:number_blocks}}
    
    \subfigure[Size frame \\ (K=(5, 5), B=4, W=512, P=32, H=256)]{
    \includegraphics[width=.30\textwidth]{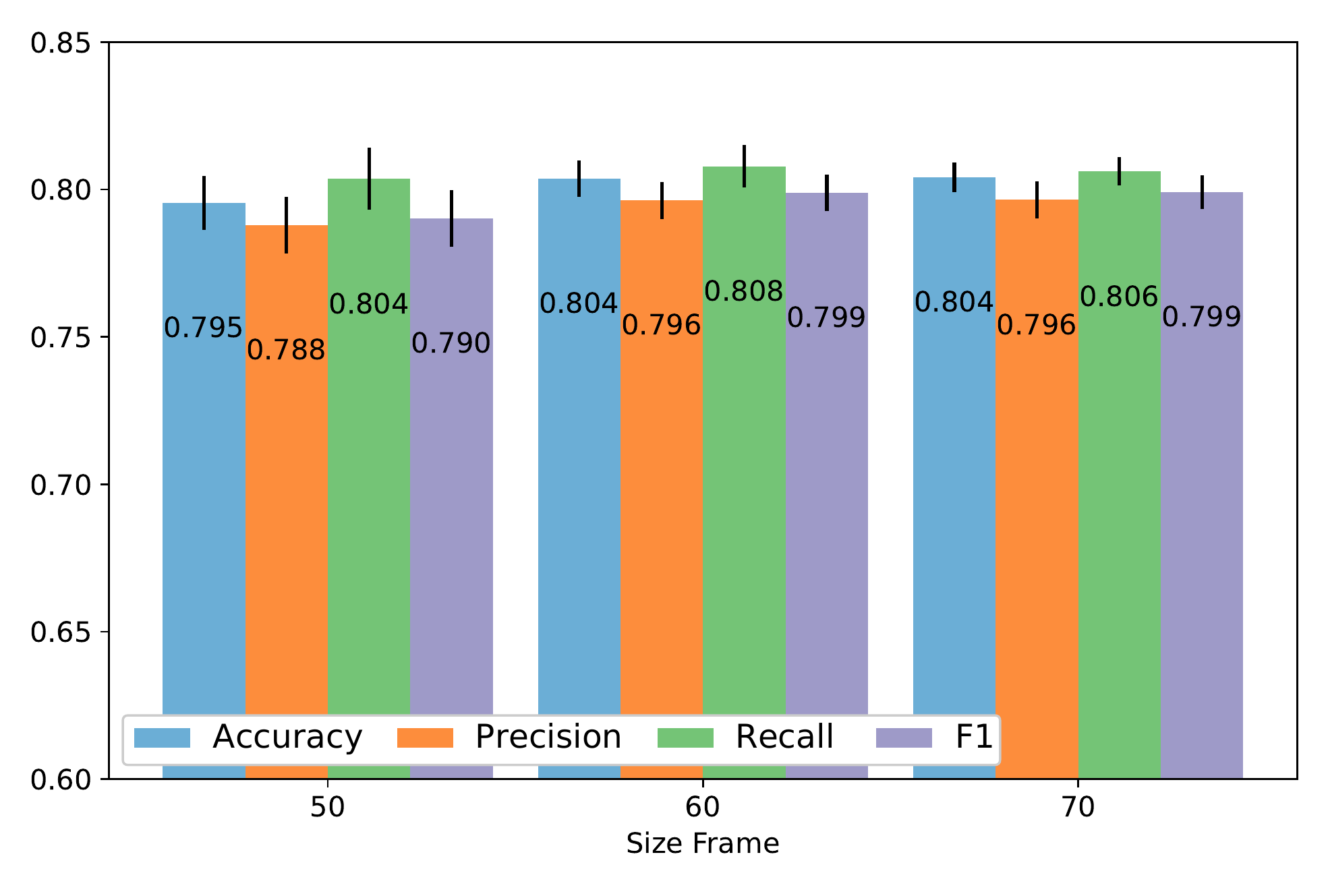}
    \label{fig:size_frame}} 
}

\centerline{
    \subfigure[Size window \\(K=(5, 5), B=4, F=60, P=32, H=256)]{
    \includegraphics[width=.30\textwidth]{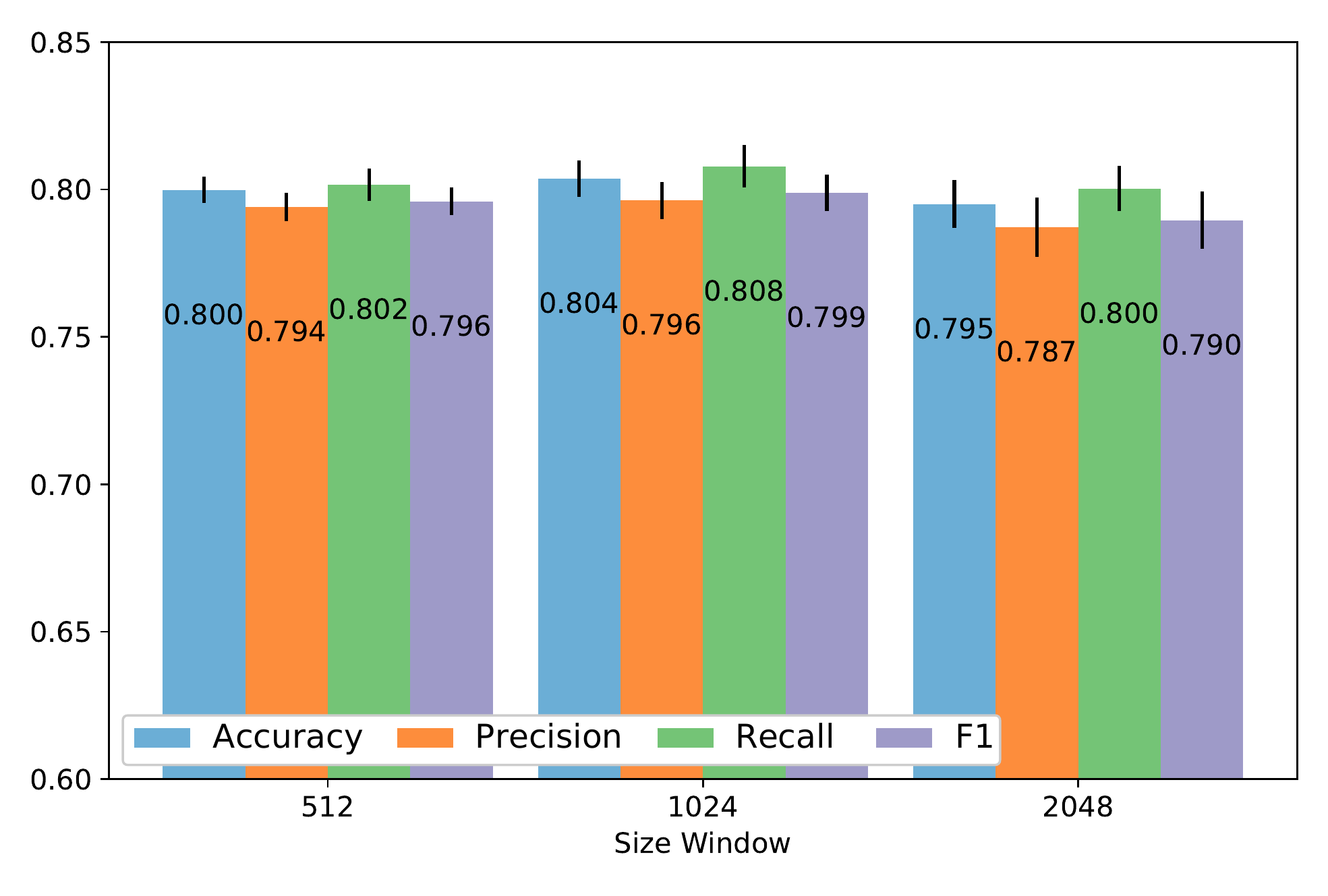}
    \label{fig:sens_threshold}} 
    \subfigure[Number Filters \\ (K=(5, 5), B=4, F=60, W=1024, H=256)]{
    \includegraphics[width=.30\textwidth]{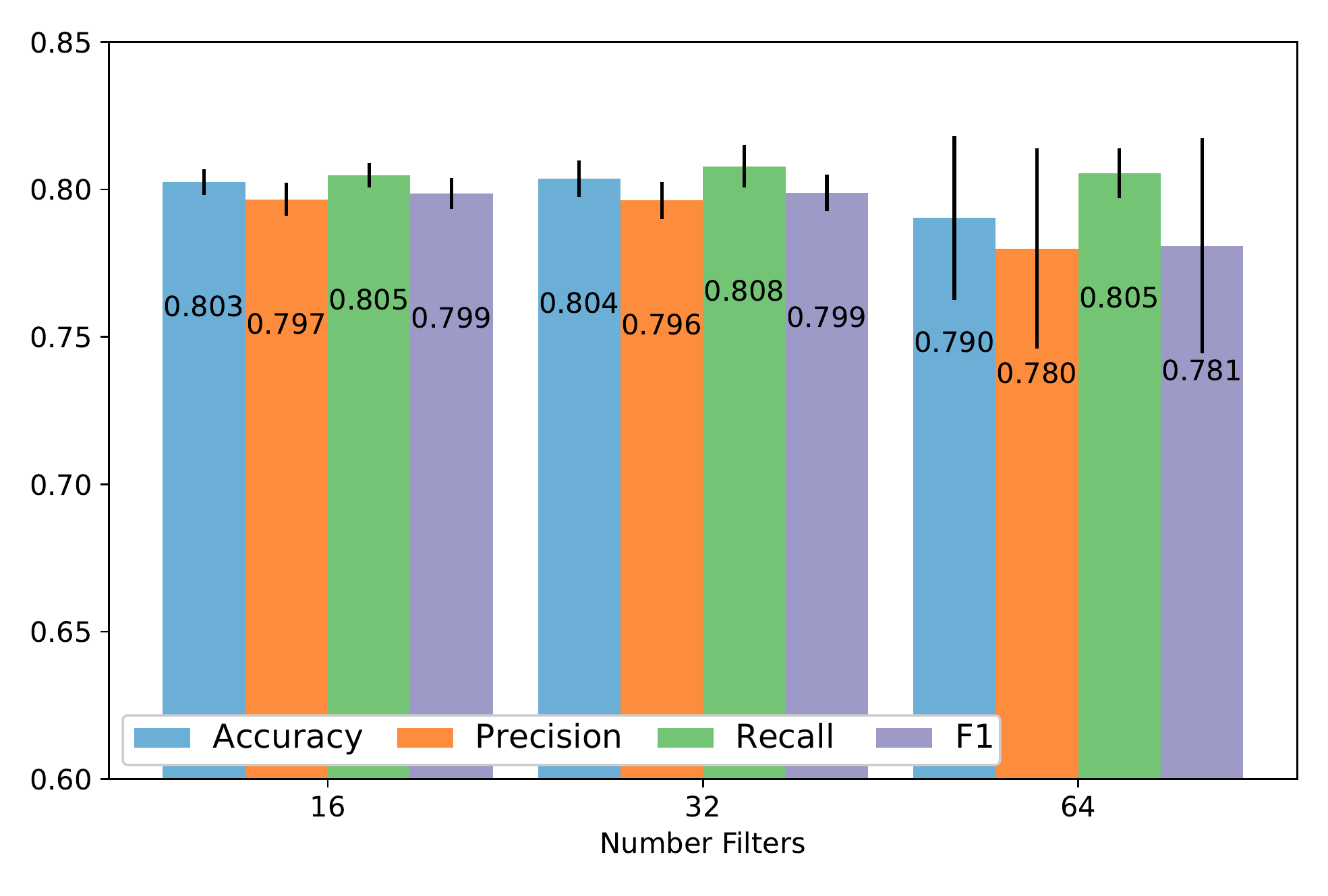}
    \label{fig:sens_topologia}} 
    
    \subfigure[Hop Length \\ (K=(5, 5), B=4, F=60, P=32, W=1024)]{
    \includegraphics[width=.30\textwidth]{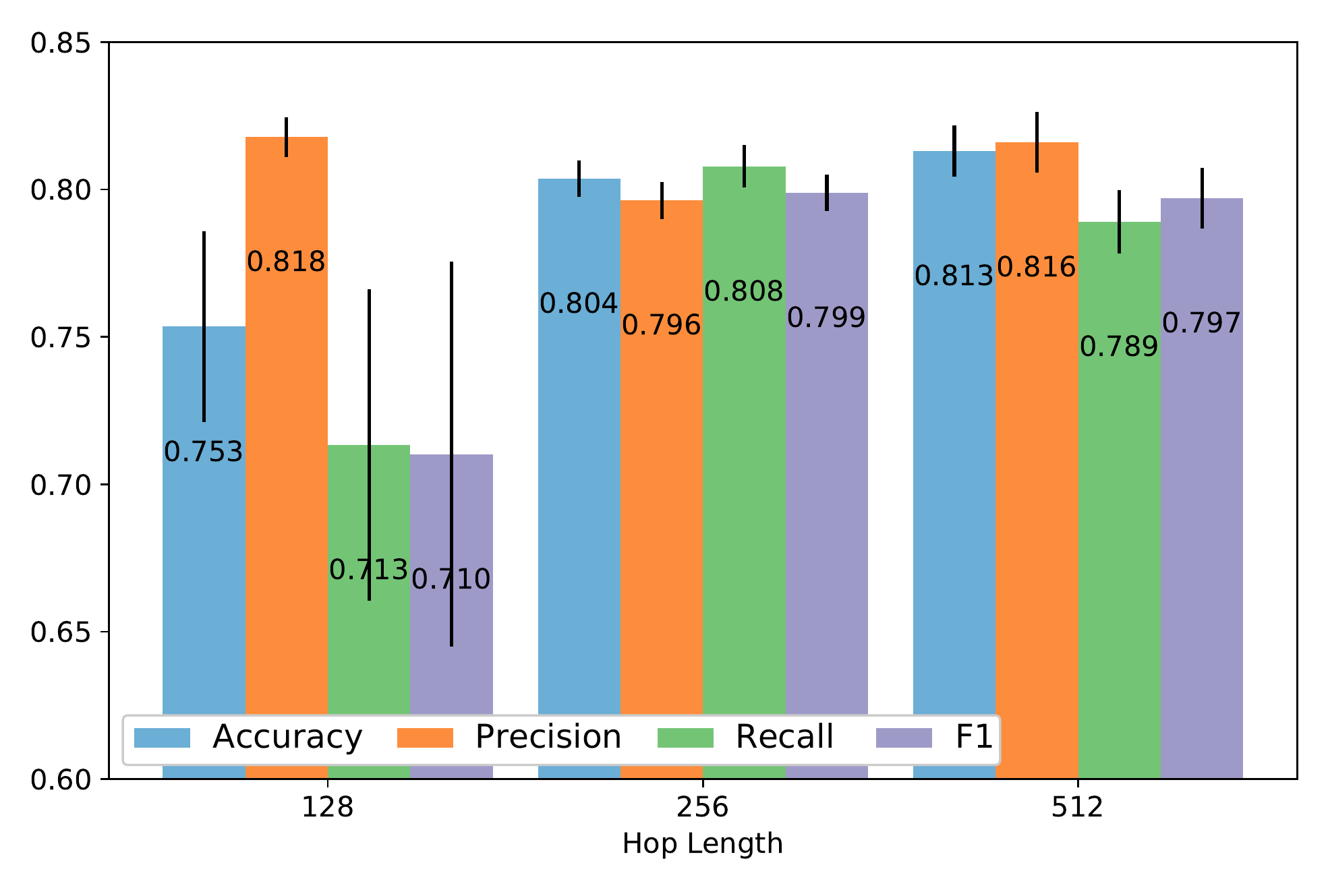}
    \label{fig:sens_hop}} 
       
} 
   \caption{Neural network sensitivity to values for multiple parameters (Dataset $=D5$) }
   \label{fig_sensi}
\end{figure}

Figure~\ref{fig_sensi} presents the result of the evaluation for six parameters: size of filters ($F$), number of blocks ($B$), size of the frame ($F$), size of the window ($W$), number of filters ($ F$) and Hop length ($L$). Each graph presents the impact of three distinct values on each of the four metrics of interest. Each bar shows the mean and standard deviation of 10 repetitions. In this study, we used only the D5 dataset as input, which we consider the most challenging.

Figure~\ref{fig:size_filters} shows the impact of varying the size of the filters ($K=\{(3.3), (5.5), (7.7))\}$ convolutional on the analyzed metrics. The results suggest that the parameter has a low impact on the overall results and that the filter size $K=(5, 5)$ has a slight advantage over the others. Figure~\ref{fig:number_blocks} shows the impact of varying the number of blocks ($B=\{3, 4, 5\}$) on the metrics and suggests that the number of blocks $B=4$ presents the best result. Figure~\ref{fig:size_frame} shows that the impact of varying frame sizes on prediction quality is small. Therefore, we considered the frame size equal to $F=60$, as it presents a more balanced result between the evaluated metrics and the recording time interval.
    The sensitivity is also small for the window size of the STFT $W$, as shown in Figure~\ref{fig:sens_threshold}), and for which we adopted the value $W=1024$.
  Figure~\ref{fig:sens_topologia} shows that the number of blocks per filter $P$ tends to generate a lower mean and a higher standard deviation impact for value $P=64$, so we adopted a lower value $P=32$.
      Finally, Figure~\ref{fig:sens_hop} shows that the hop length $H$ tends to generate a lower mean and a higher standard deviation impact for the extreme values and, therefore, we adopted the central value $H=256$.
As a result of this step, we adopted the following configurations for the rest of the evaluations: filter size $K=(5, 5)$, number of blocks $B=5$, frame size $F=60$, window size of the STFT $W=1024$, number of filters per block $P=32$ and hop length $H=256$.

\subsection{Efficacy}
\label{subsec_eva_efficacy}

We divided the evaluation of the efficacy of our model into two stages. First, we consider the results predicted by the model for each of the six available datasets. In the second, we compare our model with the current state of the art considering our most challenging scenario ($D5$) and the dataset used in the referred work ($D6$).

\begin{figure}[htb]
\centerline{
    \includegraphics[width=1.0\textwidth]{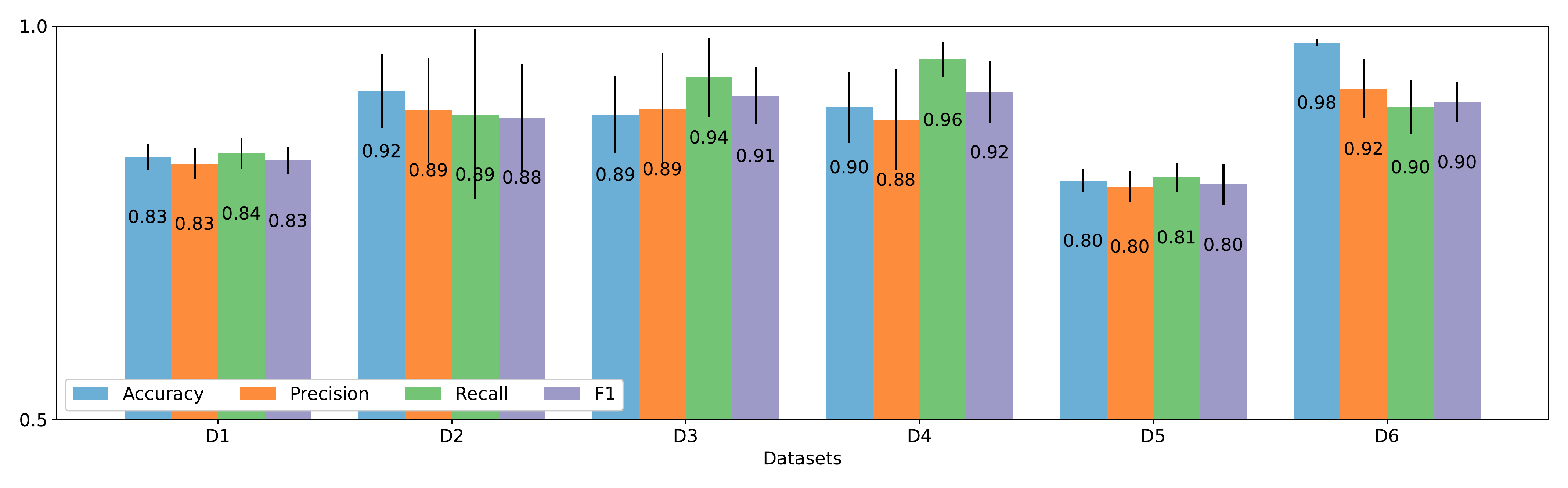}
}
\caption{Performance of the proposed solution considering multiple datasets ($D=\{D1, D2, D3, D4, D5, D6\}$)}
\label{fig:efficacy_datasets}
\end{figure}

Figure~\ref{fig:efficacy_datasets} shows the results obtained for all six datasets. We display four bars for each dataset: one bar for each metric of interest showing the respective means and standard deviation for 10 executions. It is possible to observe that in all evaluated scenarios, our model obtained a score greater than 80\% for each of the considered metrics. Two other important observations concern the degree of complexity of each evaluated dataset. First, datasets D1 and D5 have average scores 9.25\% lower than the other datasets. One of the hypotheses for this behavior is that male \aedes samples present in both datasets are considerably more difficult to be classified than the others. Second, it seems equally challenging to distinguish between different groups of mosquitoes and ambient noise samples. Our main hypothesis for this behavior is the presence of ambient noise in samples positive for mosquitoes. This situation could cause the prediction model not to be able to distinguish between positive samples and negative samples. In particular, this last observation is important because in real application scenarios, for example, within smart mosquito traps, ambient noise could reduce the classifier's efficiency.

\begin{figure}[htb]
\centerline{

    \subfigure[Dataset D1]{
    \includegraphics[width=.35\textwidth]{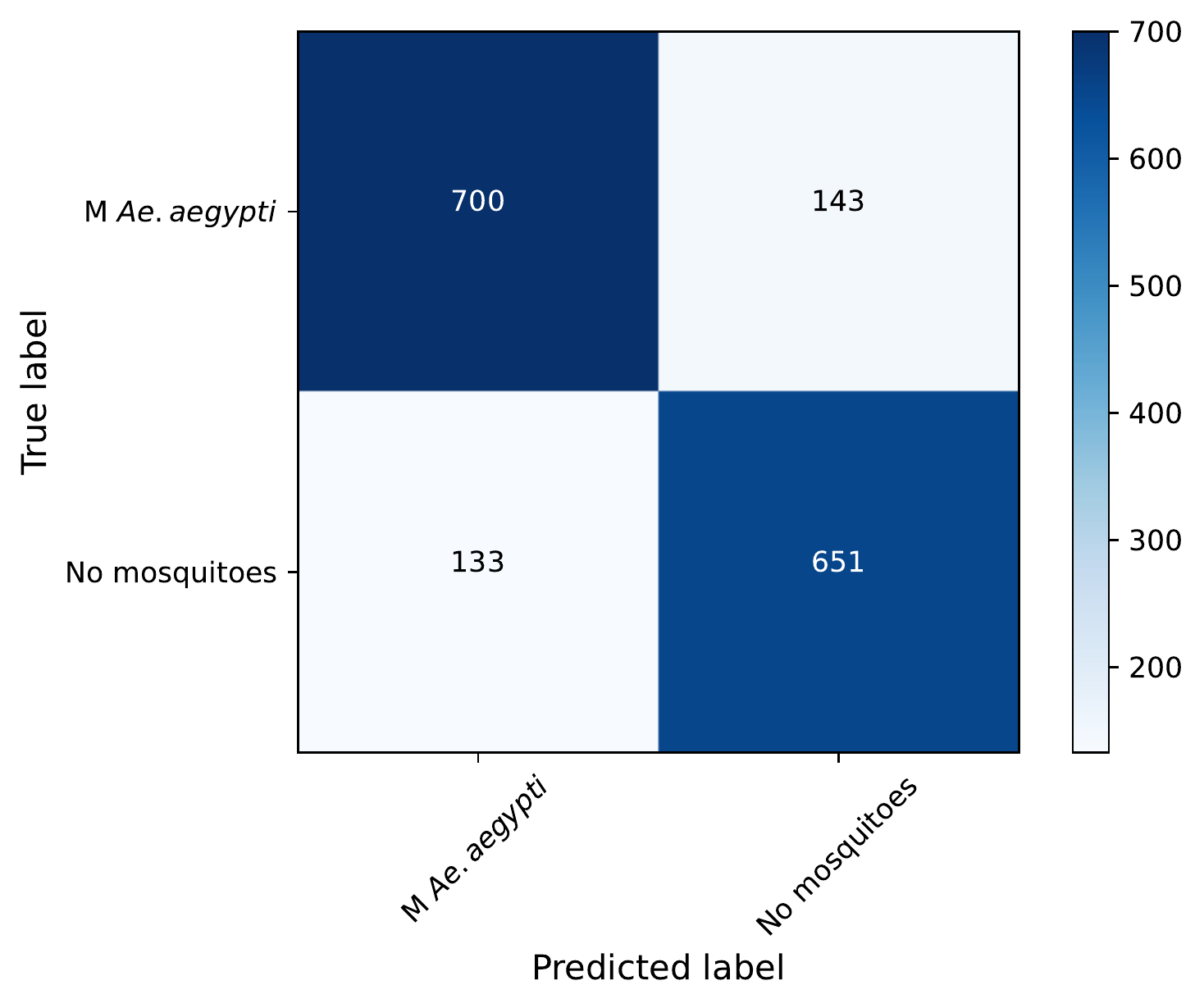}
    \label{Confusion_matrix_D1}} 
    
    \subfigure[Dataset D2]{
    \includegraphics[width=.35\textwidth]{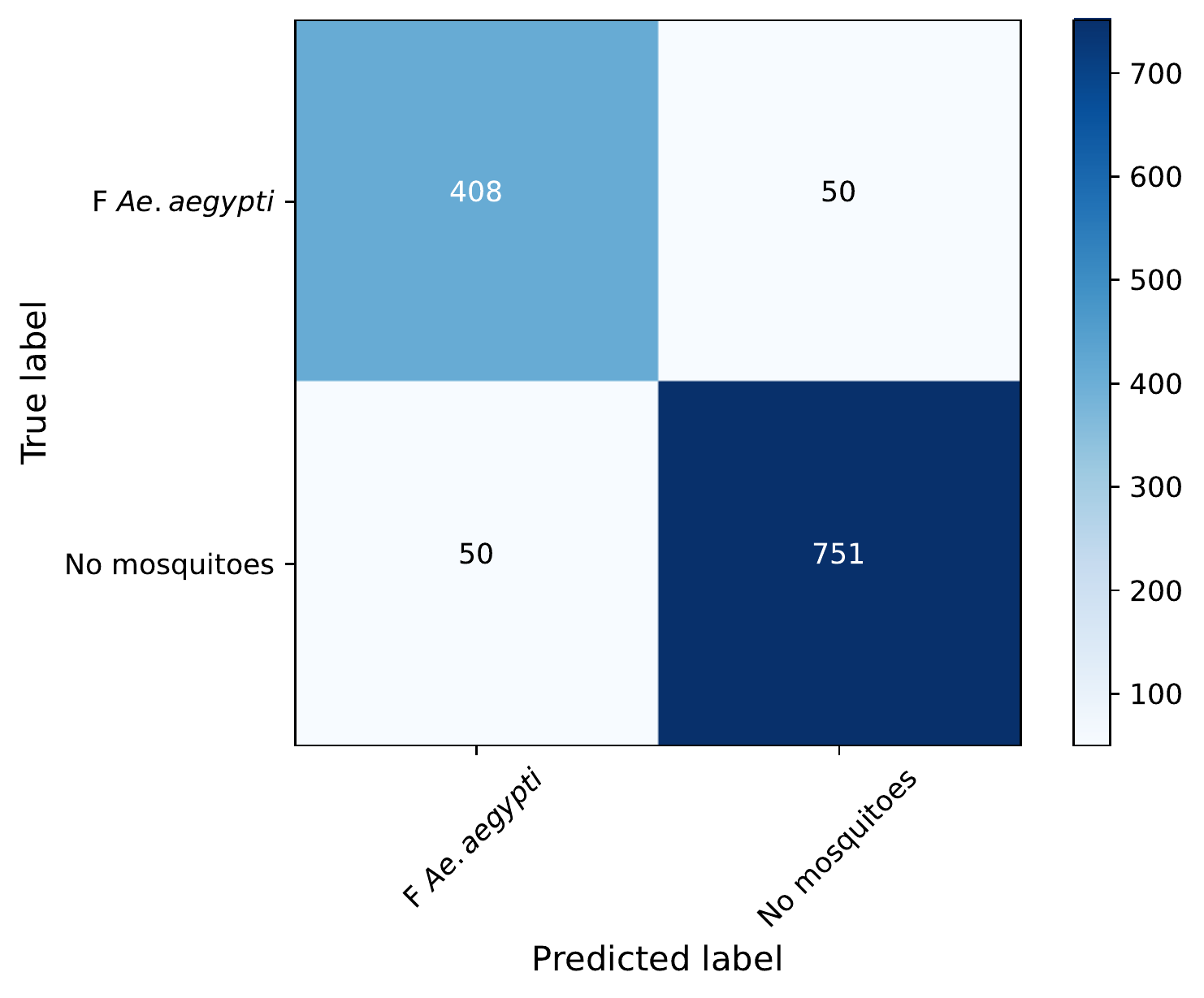} \label{fig:Confusion_matrix_D2}}
    
    \subfigure[Dataset D3]{
    \includegraphics[width=.35\textwidth]{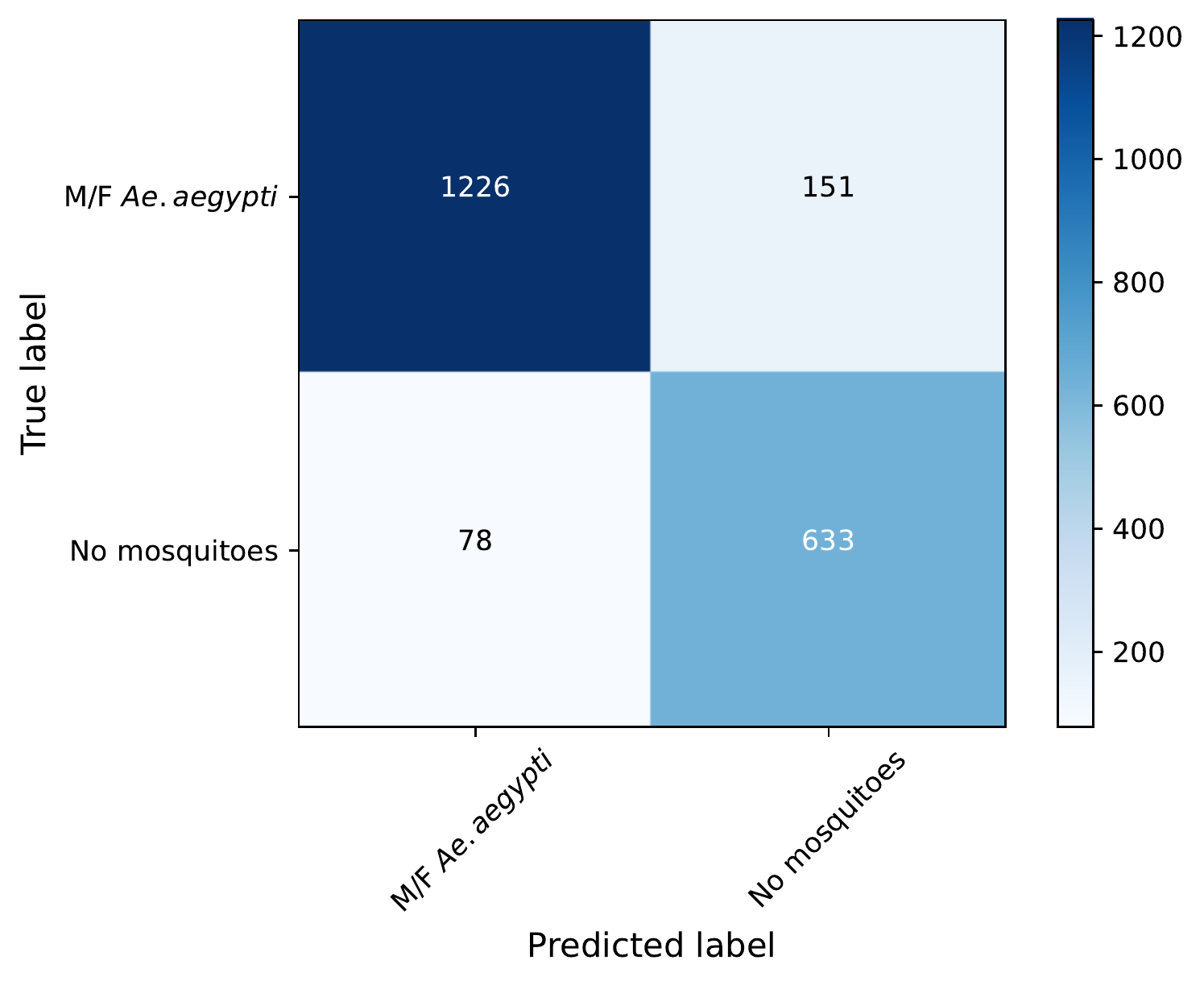}
    \label{fig:Confusion_matrix_D3}} 
} 

\centerline{

    \subfigure[Dataset D4]{
    \includegraphics[width=.35\textwidth]{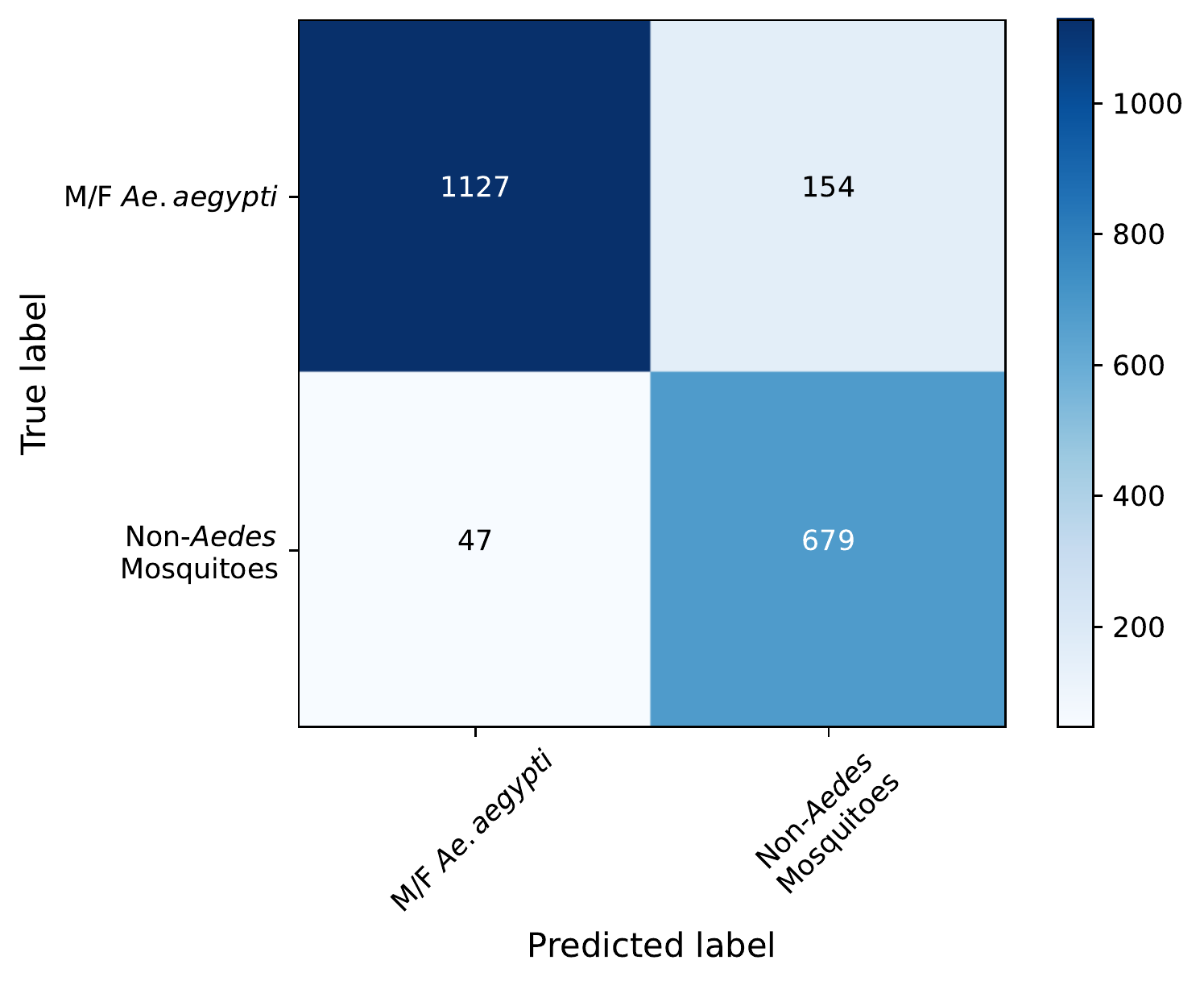}
    \label{Confusion_matrix_D4}} 
    
    \subfigure[Dataset D5]{
    \includegraphics[width=.35\textwidth]{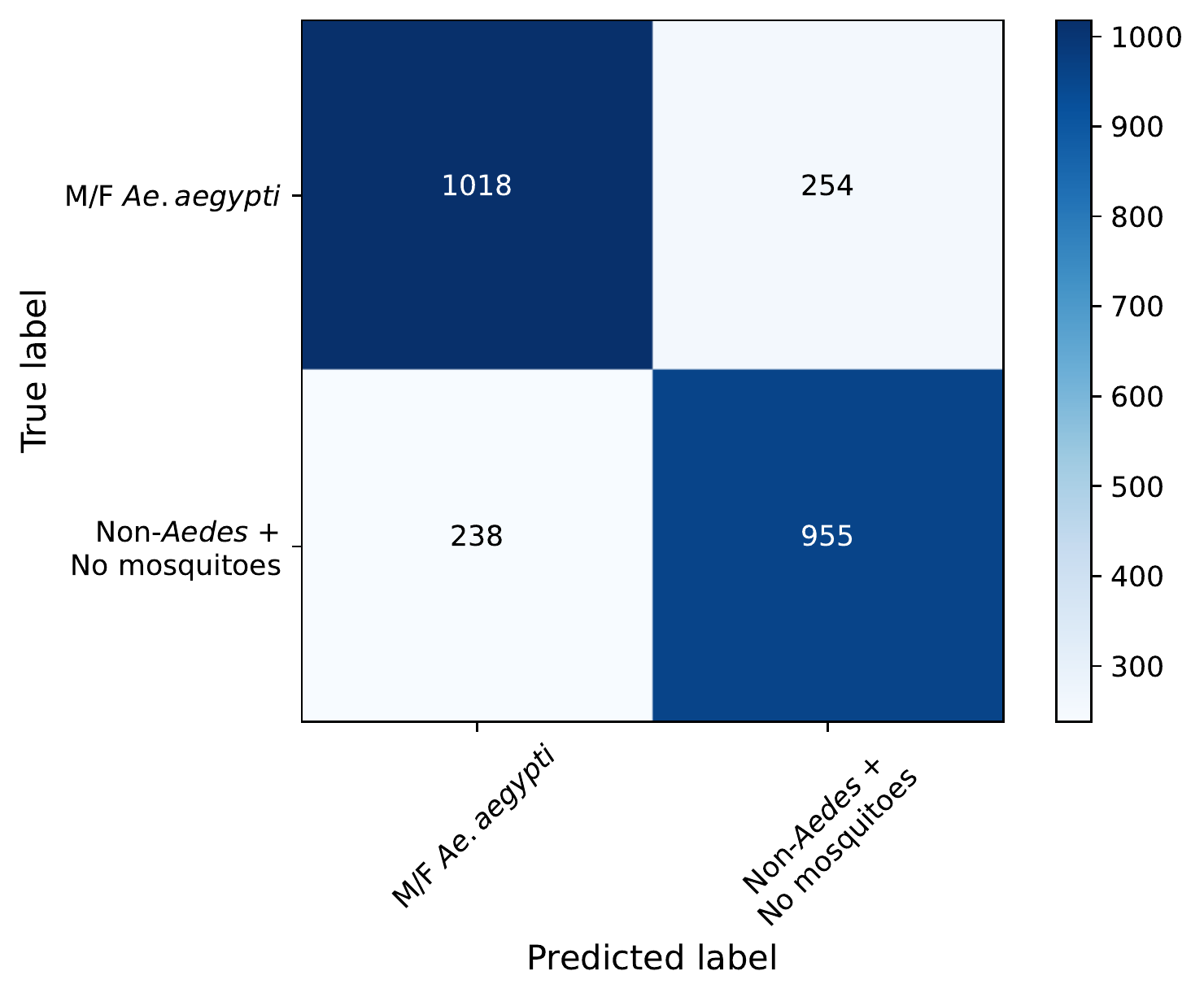} \label{fig:Confusion_matrix_D5}}
    
    \subfigure[Dataset D6]{
    \includegraphics[width=.35\textwidth]{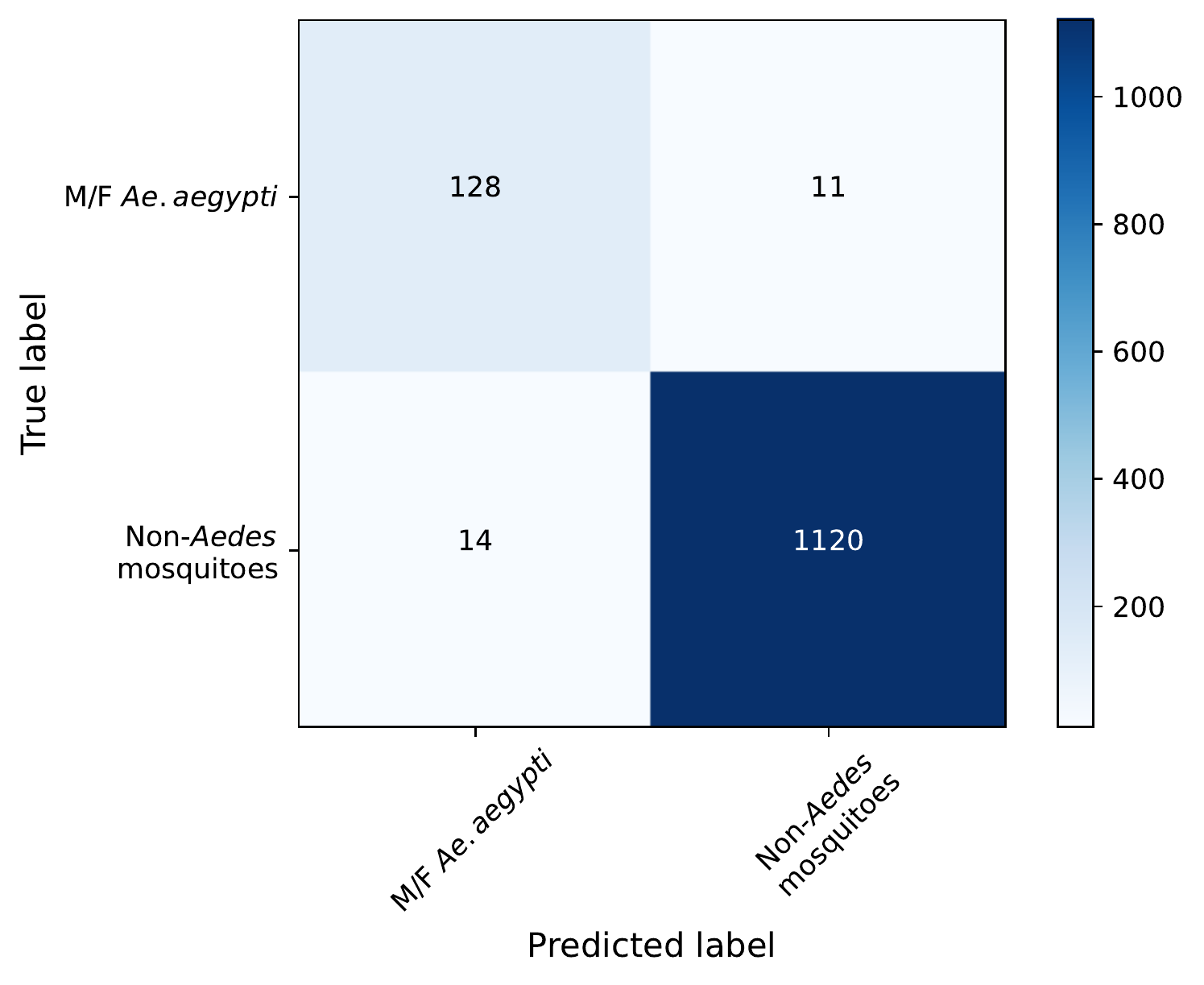}
    \label{fig:Confusion_matrix_D6}} 
}

\caption{Detailed performance of the proposed solution considering multiple datasets ($D=\{D1, D2, D3, D4, D5, D6\}$)}
\label{fig:matrix}
\end{figure}

In Figure \ref{fig:matrix}, we present the confusion matrices obtained for each of the six datasets. The display of records is performed in terms of their current classes and their predicted classes. True labels are arranged in rows, and predicted labels are in columns. Each cell represents the average of samples sorted during the cross-validation process. The presence of higher values on the matrix's diagonal demonstrates the model's predictive capacity, as they add up to represent the total number of hits in the model. It is possible to verify that in all the evaluated scenarios, the values on the diagonal are higher concerning the others -- this is a strong indication of the high predictive capacity of the proposed model. As another observation, we did not find evidence of shortcomings in classifying that could indicate a preference for a specific label.

\begin{figure}[htb]
\centerline{
    \includegraphics[width=0.70\textwidth]{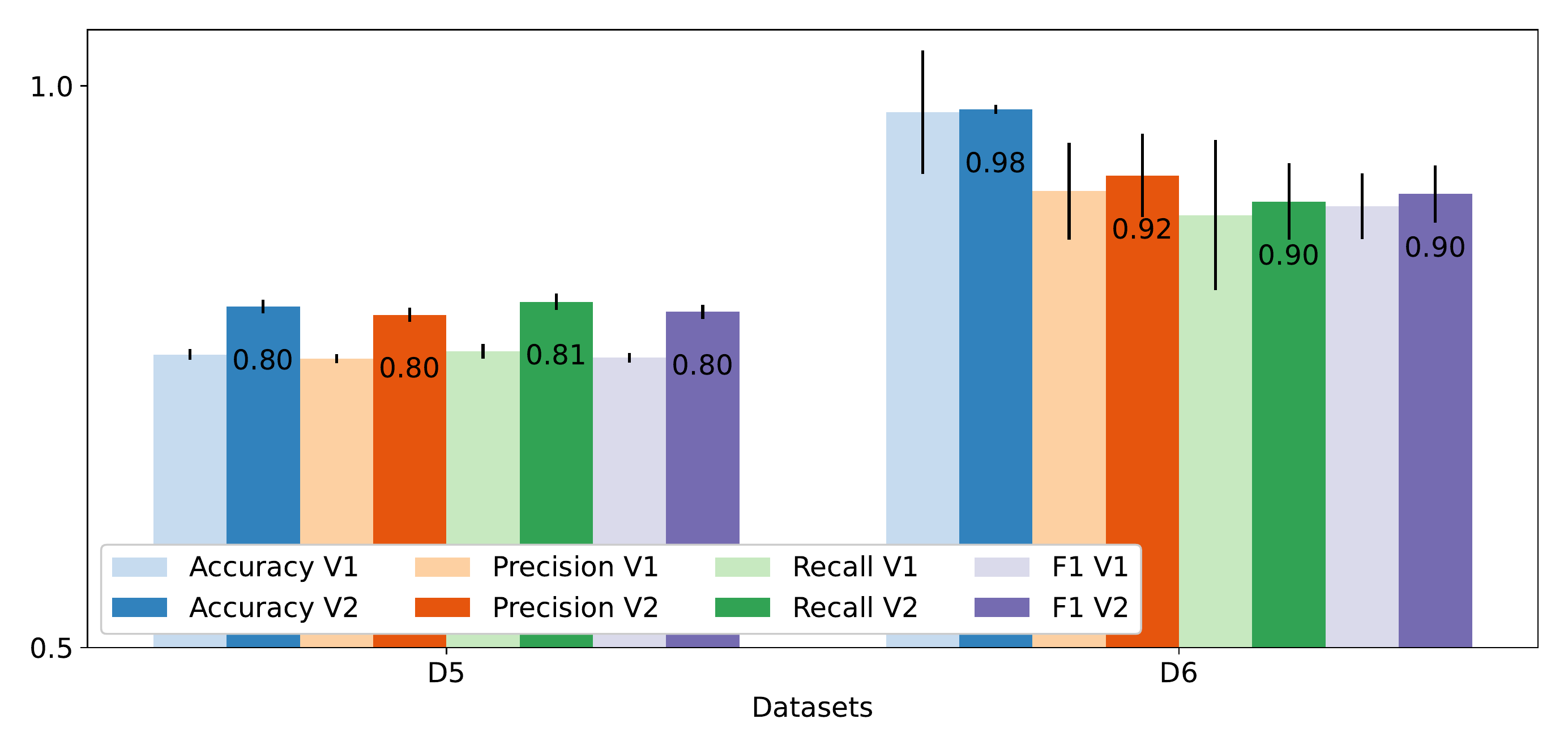}
    
}
\caption{Comparison with the state-of-the-art ($D=\{D5, D6\}$)}
\label{fig:comparision_models}
\end{figure}

Figure~\ref{fig:comparision_models} presents the results obtained with datasets D5 and D6 for the two neural networks (current and prior). For each dataset, eight bars are presented, representing the performance metrics for each neural network. 
Like the other plots above, each bar represents the mean and standard deviation obtained through cross-validation after ten runs.

From Figure~\ref{fig:comparision_models}, we can observe that in both datasets and in all metrics, our current algorithm performed better than the state-of-the-art. In the comparison using the D5 dataset, our algorithm presented an average score of 4.16\% higher than the original algorithm. In the comparison using the D6 dataset, our algorithm presented a lower average score, approximately 1\%. A final observation is that the D5 dataset appears more challenging than the D6 dataset.

\subsection{Proof of Concept: Testing the Residual CNN in a Smartphone App}
\label{subsec_eva_proof}

As a proof of concept, we developed an Android-based smartphone app for  identifying \aedes mosquitoes in a semi-controlled environment. From our experiments, the app has shown to meet the necessary performance requirements for popular off-the-shelf smartphones as it does not require substantial processing power and has a minimum OS requirement for Android Oreo 8.1 (as of today, the current version is 13). The app's interface has three components: (i) a frame that displays the spectrogram (STFT) resulting from the decomposition of the recorded audio; (ii) a sound intensity identifier frame; and (iii) a number on a percentage scale identifying the probability that the sample contains signs of an \aedes mosquito.

We wrote the smartphone app in Java (v8.0) and C++ (v11) and built the user interface on top of standard Android libraries. We adopted the following techniques to reduce the model prediction time and save battery. First, we implemented the mathematical operations of the Fourier transform in C++ and compiled them for the ARM x64 architecture. We trained the neural network using cloud computing and executed it in the smartphone through the TensorFlow Lite library (v2.1.0), which implements low-level functions.

We developed the neural network to have the minimum possible number of parameters to reduce computing and energy costs. In total, the developed model has about  1 million (927,458) trainable parameters. This value is 18.5\% smaller than the previous network, which had about 1.2 million (1,137,858) parameters. The value of our current network is also considered low compared to models such as ResNet152 \cite{residual_2016}, which has approximately 60 million parameters, and DenseNet201 \cite{denseNet}, which has approximately 20 million parameters.

We also adopted optimizations related to the model and the interpreter to run the app on popular smartphones. During the training stage, the model underwent a dynamic range quantization process. This technique reduced the model size by 18.5\% and reduced the inference time by approximately 2 times, keeping the model's effectiveness virtually unchanged. Regarding the interpreter, we adopted two optimization strategies. First, we used the multiple processing threads option, which reduced the inference time. Second, we used an inference pipeline that performs the feature extraction flow, classification, and results display independently.

We ran the app on an Asus Zenfone Max M3 Qualcomm Snapdragon 430 64-bit Octa core 4G RAM Android 8.1 and a Motorola Moto E6 Play Cortex-A53 1.5 GHz Quad Core 2 GB RAM Android 9.0 smartphones, both entry-level models in Brazil at the time of release. The average time for each inference was approximately 320 ms.

\begin{figure}[htb]

\centerline{
    \subfigure[]{
    \includegraphics[width=.21\textwidth]{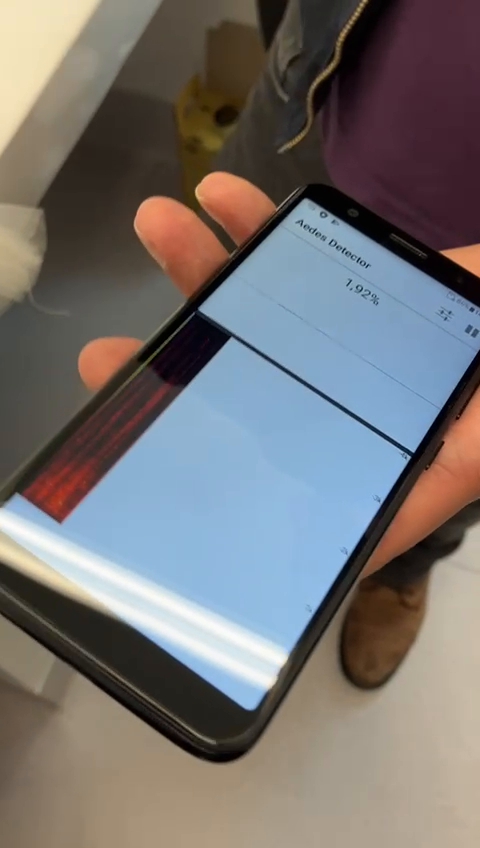}
    \label{fig:test_lab_aedes_aegypti_male}} 
                
    \subfigure[]{
    \includegraphics[width=.21\textwidth]{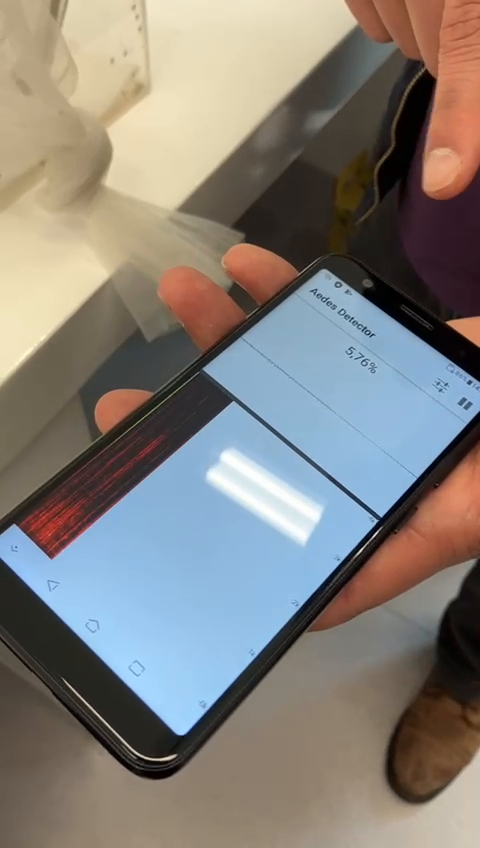} \label{fig:test_lab_aedes_aegypti_female}} 
    
    \subfigure[]{
    \includegraphics[width=.21\textwidth]{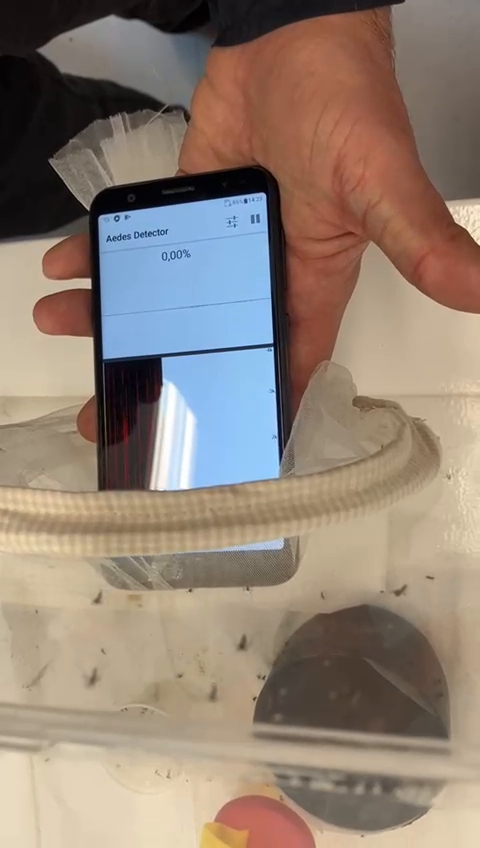}
    \label{fig:test_lab_aedes_albipictus_male}} 

    \subfigure[]{
\includegraphics[width=.21\textwidth]{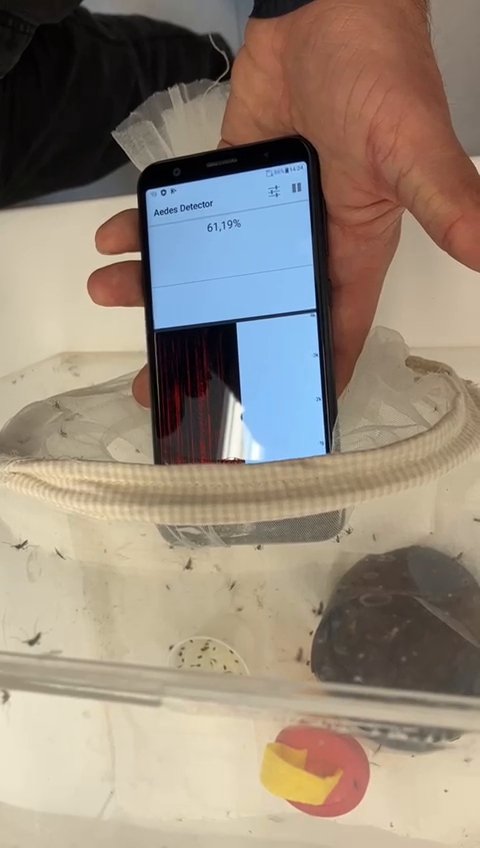} \label{fig:test_lab_aedes_albopictus_female}}    
}

\centerline{
    \subfigure[]{
    \includegraphics[width=.21\textwidth]{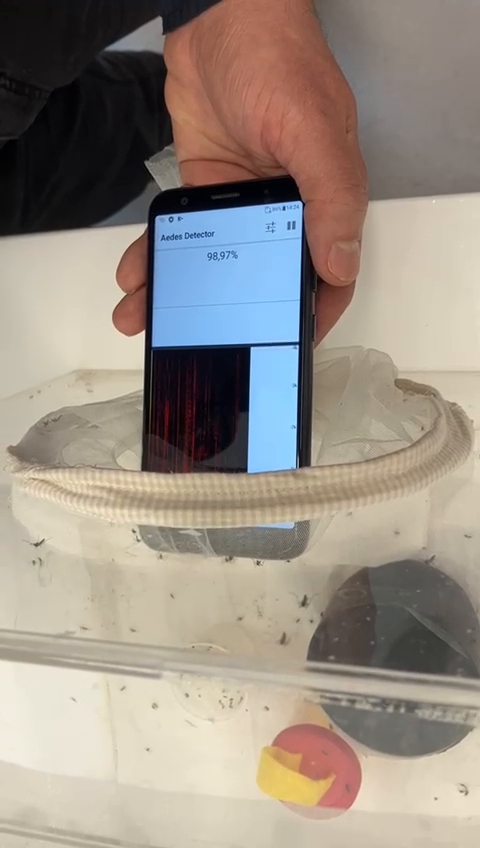}
    \label{fig:test_lab_aedes_sierrensis}} 
    
    \subfigure[]{
    \includegraphics[width=.21\textwidth]{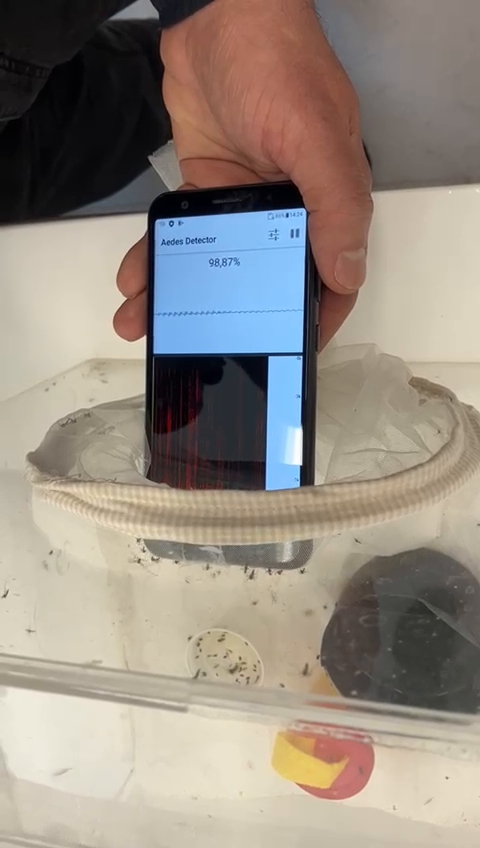} \label{fig:test_lab_anopheles_quadrimaculatus}} 
   
    \subfigure[]{
    \includegraphics[width=.21\textwidth]{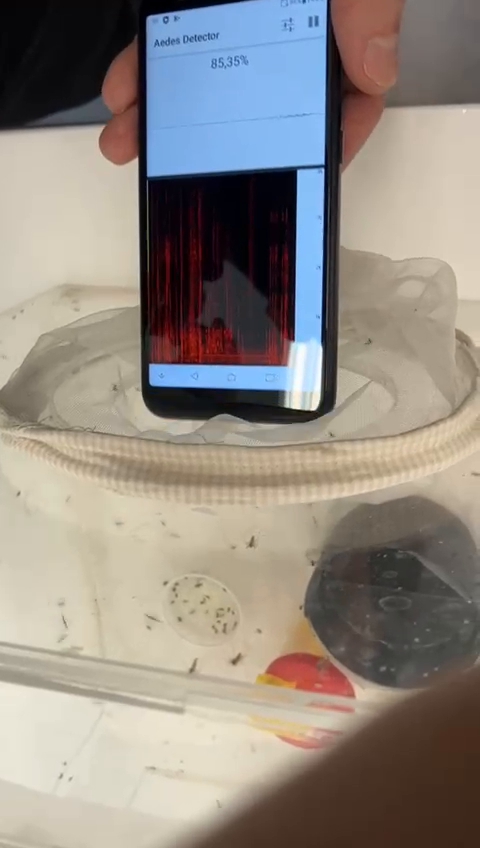}
    \label{fig:test_lab_anopheles_stephensi}} 

    \subfigure[]{
    \includegraphics[width=.21\textwidth]{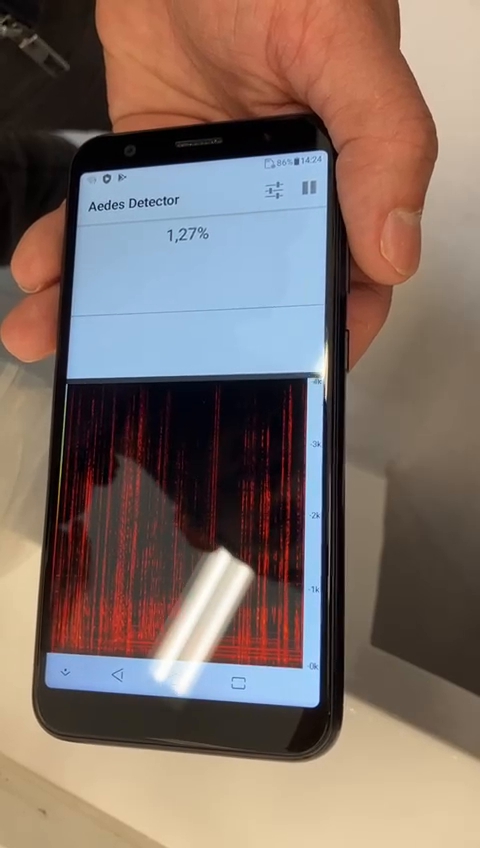}
    \label{fig:test_lab_environment_noise}}      
}
\caption{
Sequence of images of a test with mosquitoes carried out in the laboratory}
\label{fig:lab_test}
\end{figure}

We consider that in an environment with semi-controlled noise (a biology laboratory without acoustic isolation), the model presented high stability to identify the presence of \aedes mosquito effectively. 
This fact is illustrated by Figure~\ref{fig:lab_test}, which presents a sequence of 8 images\footnote{A full video version of the experiment, along with the audio recordings of \aedes mosquitoes and other paper artifacts, can be obtained at \url{https://github.com/ComputerNetworks-UFRGS/Aedes_Detector_Smartphone_App_Paper}.} captured during a laboratory test of the application with \aedes mosquitoes. The sequence shows:
(a) Recording of ambient noise with the prediction of low probability of \aedes mosquito; 
(b) Increased ambient noise level with low probability prediction of \aedes mosquito; 
(c) Smartphone closes to the cage containing \aedes mosquitoes; 
(d) Identification of increased probability of predicted \aedes mosquito; 
(e) High probability for predicted \aedes class; 
(f) High probability continuity for predicted \aedes class; 
(g) Removing the smartphone from the mosquito cage and consequent reduction in the probability; 
(h) Complete removal of the mosquito cage and sudden reduction of probability for \aedes mosquitoes. 

\section{Final Considerations}
\label{sec:conclusion}

Using a device that is universally present in modern society, such as the smartphone, can be an important weapon in tackling the proliferation of \aedes mosquitoes. However, we need to overcome several practical challenges, including the limitation of existing datasets and computational constraints of devices used by the target audience.

In this work, we: (i) present a new dataset, (ii) propose a neural network topology, (iii) develop a training method as steps towards the implementation of monitoring applications of \aedes mosquitoes on low-cost mobile devices. The results of an evaluation with unpublished and pre-existing real datasets show the advance provided by our proposal towards state-of-the-art. The proposed neural network is 18.5\% smaller in terms of the number of parameters, which is the main computational efficiency factor of the technique. The lean neural network combined with the improved training technique can generate results equivalent to or slightly superior to state-of-the-art in terms of accuracy, precision, recall, and F1 in all considered scenarios. In addition, implementing a functional prototype demonstrates the feasibility of running (without failures and crashes) the proposed solution on popular smartphones.

As future work, we envision three main research directions: expanding the study to other devices, expanding the dataset, expanding the comparison with other state-of-the-art works, evaluating the execution in even more limited devices such as IoT, advancing the development of neural networks to update with advances in the area, develop techniques to make the neural network resistant to noise generated deliberately by malicious users attempting to induce false positives.


\section*{Acknowledgements}

We thank Prof. Dr. Gonçalo Nuno Corte Real Ferraz de Oliveira (UFRGS, Brazil) for his contributions to this study. This study was financed in part by the Coordenação de Aperfeiçoamento de Pessoal de Nível Superior - Brasil (CAPES) - Finance Code 001. We also received funding from Rio Grande do Sul Research Foundation (FAPERGS) (grants 22/2551-0000841-0, 22/2551-0000390-7, and 22/2551-0000603-5) and NVIDIA -- Academic Hardware Grant.

\bibliographystyle{cas-model2-names}
\bibliography{paper}

\end{document}